\documentclass[pdflatex,sn-mathphys-num]{sn-jnl}

\usepackage{graphicx}%
\usepackage{multirow}%
\usepackage{amsmath,amssymb,amsfonts}%
\usepackage{amsthm}%
\usepackage{mathrsfs}%
\usepackage[title]{appendix}%
\usepackage{xcolor}%
\usepackage{textcomp}%
\usepackage{manyfoot}%
\usepackage{booktabs}%
\usepackage{algorithm}%
\usepackage{algorithmicx}%
\usepackage{algpseudocode}%
\usepackage{listings}%
\usepackage{comment}
\usepackage{lineno}

\newcommand{\bfx}{{\boldsymbol x}}%
\newcommand{\bfy}{{\boldsymbol y}}%
\newcommand{\zz}{{\mathbb Z}}%
\newcommand{\nn}{{\mathbb N}}%
\newcommand{\pp}{{\mathbb P}}%

\newcommand{\bfzero}{{\boldsymbol 0}}%
\theoremstyle{thmstyleone}%
\newtheorem{theorem}{Theorem}%
\newtheorem{corollary}{Corollary}
\newtheorem{lemma}{Lemma}

\newtheorem{proposition}[theorem]{Proposition}%

\theoremstyle{thmstyletwo}%
\newtheorem{example}{Example}%
\newtheorem{remark}{Remark}%

\theoremstyle{thmstylethree}%

\begin{document}

\title[Social networks as intersections graphs of random walks]{Animal social networks as intersections graphs of random walks}


\author{\fnm{Paolo} \sur{Cermelli}}\email{paolo.cermelli@unito.it}

\author{\fnm{Silvia} \sur{Marchese}}\email{silviamarchese98@gmail.com}

\author{\fnm{Laura} \sur{Sacerdote}}\email{laura.sacerdote@unito.it}
\author{\fnm{Cristina} \sur{Zucca}}\email{cristina.zucca@unito.it}

\affil{\orgdiv{Department of Mathematics}, \orgname{University of Turin}, \orgaddress{\street{Via Carlo Alberto, 10}, \city{Torino}, \postcode{10123}, \country{Italy}}}


\abstract{We study here the social network generated by the asynchronous visits, to a fixed set of sites, of mobile agents modelled as independent random walks on the plane lattice. The social network is constructed by assuming that a group of agents are associated if they have visited the same set of sites within a finite time interval. This construction is an instance of a random intersection graph, and has been used in the literature to study association networks  
in a number of animal species. We characterize the mathematical structure of these networks, which we view as one-mode projections of suitable bipartite graphs or, equivalently, as 2-sections of the corresponding hypergraphs. We determine analytically the probability distribution of the random bipartite graphs and hypergraphs associated to this construction, and suggest that association networks generated by the use of common resources are better described by hypergraphs rather than simple projected graphs, that miss important information regarding the actual associations among the agents.}




\maketitle

\section{Introduction}

Animal social networks represent an active area of research in the study of animal behavior \cite{krause},\cite{bohr}.  Broadly speaking, these can be grouped in two families: interaction and association networks. We are interested here in association networks, defined either through spatial proximity of the individuals, or through (possibly asynchronous) visits to the same sites, for instance, foraging \cite{foraging} or reproduction sites \cite{tortoise,parasite}.  Such models may be viewed as instances of a general class of association networks referred to as Gambit of the Group models \cite{james}, in which individuals are associated if they belong to a common group, defined  in our case by the visits to a shared site.

Animal association networks are useful for many reasons: for instance, they account for the role of the environment on the structure of animal social networks and yield insights on the factors that determine the choice of refuges or foraging sites. Also, they may be the only tool available to determine the social structure of solitary species. Further, they can be used to study the transmission of diseases or parasite infections that do not necessarily involve direct contacts between infected and susceptible individuals, but are transmitted by a host species that resides in foraging sites or refuges \cite{silk}. For example, in \cite{parasite} the authors show that the spread of parasites in the lizard Egernia stokesii is better explained by the association network derived from the use of common refuges, rather than by the social structure proper of the group.  

\begin{figure}[ht]
	\begin{center}
 \includegraphics[width=0.23\textwidth]{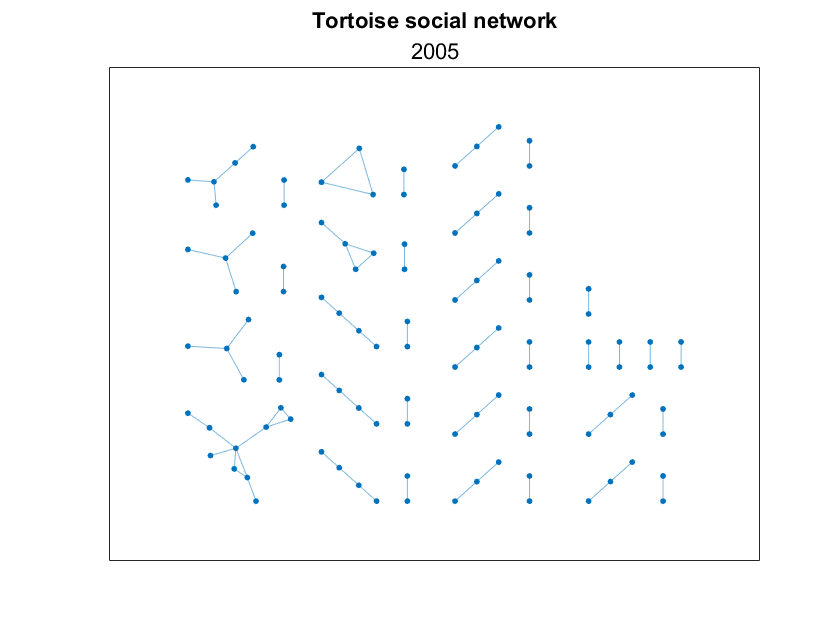}
  \includegraphics[width=0.23\textwidth]{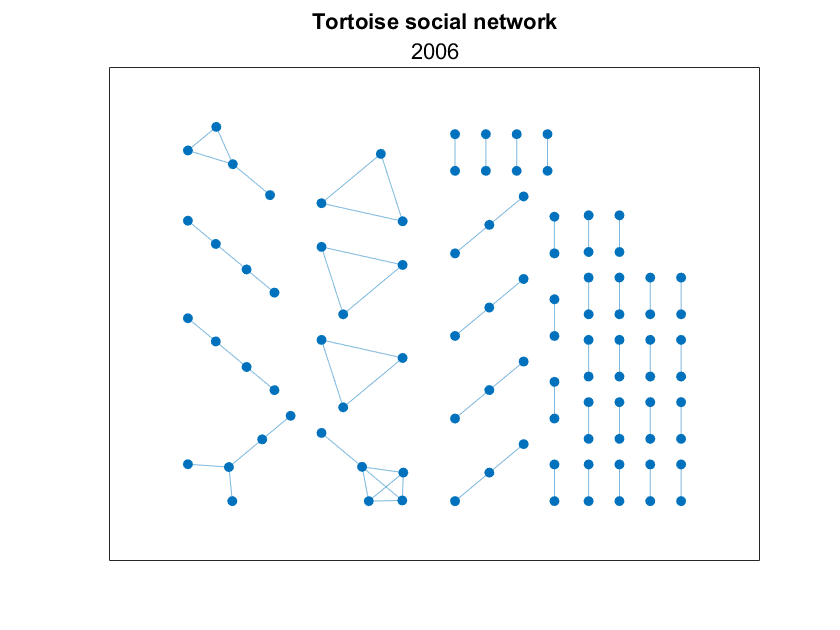}
   \includegraphics[width=0.23\textwidth]{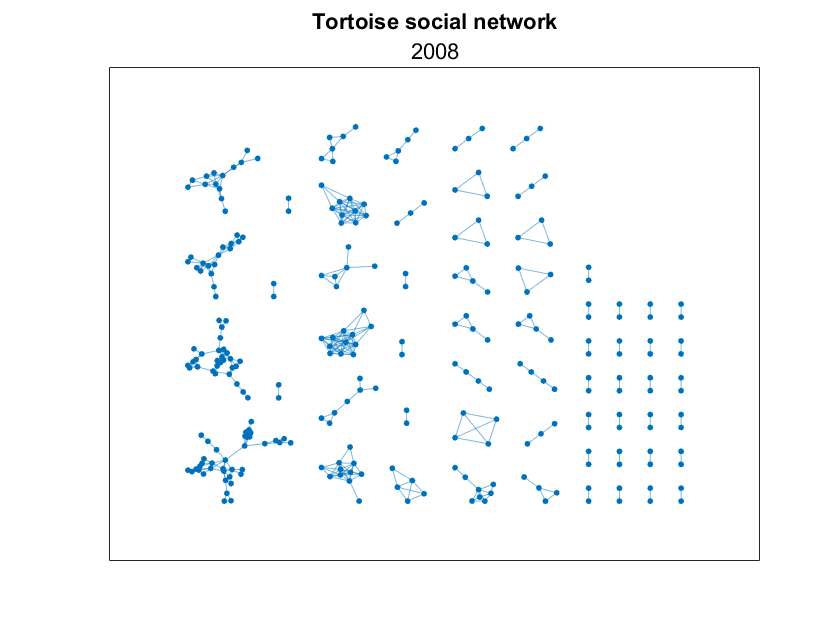}
  \includegraphics[width=0.23\textwidth]{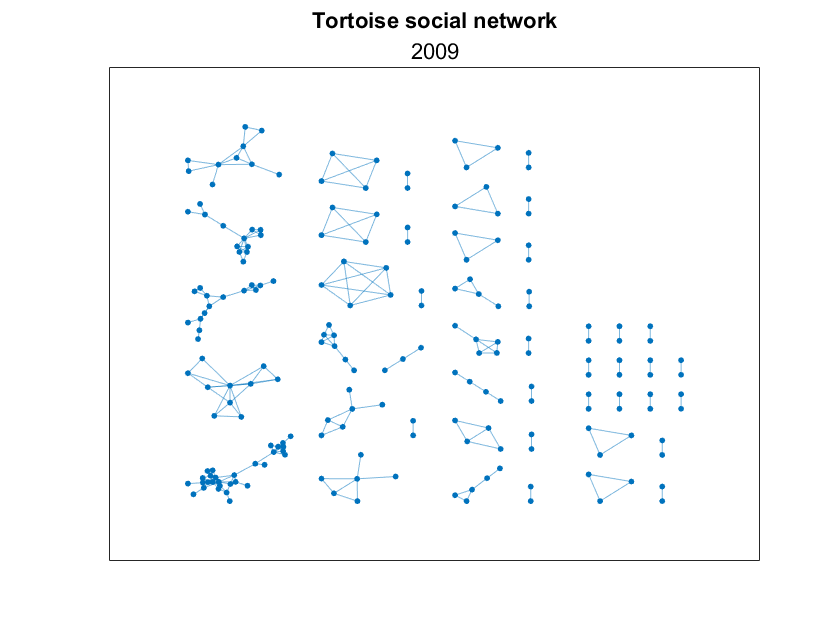}

  \includegraphics[width=0.23\textwidth]{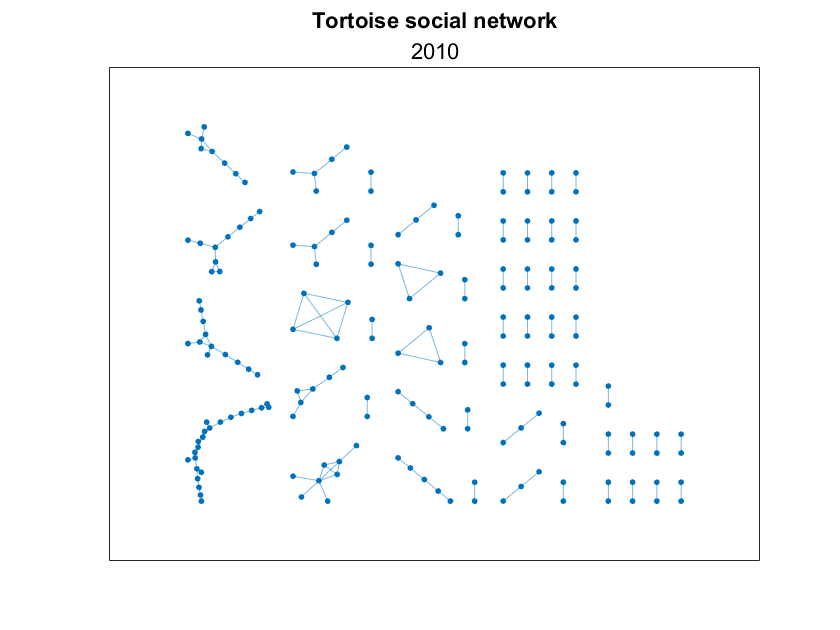}
   \includegraphics[width=0.23\textwidth]{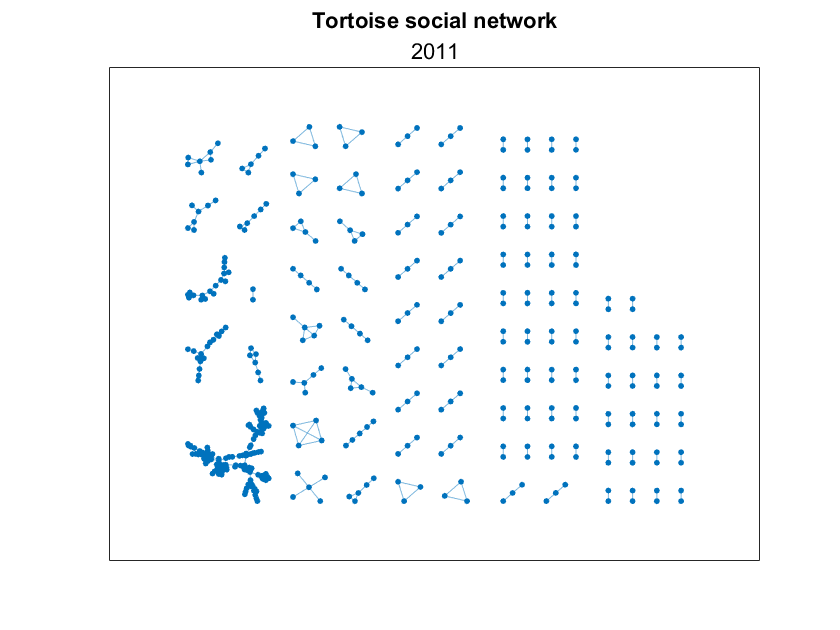}
  \includegraphics[width=0.23\textwidth]{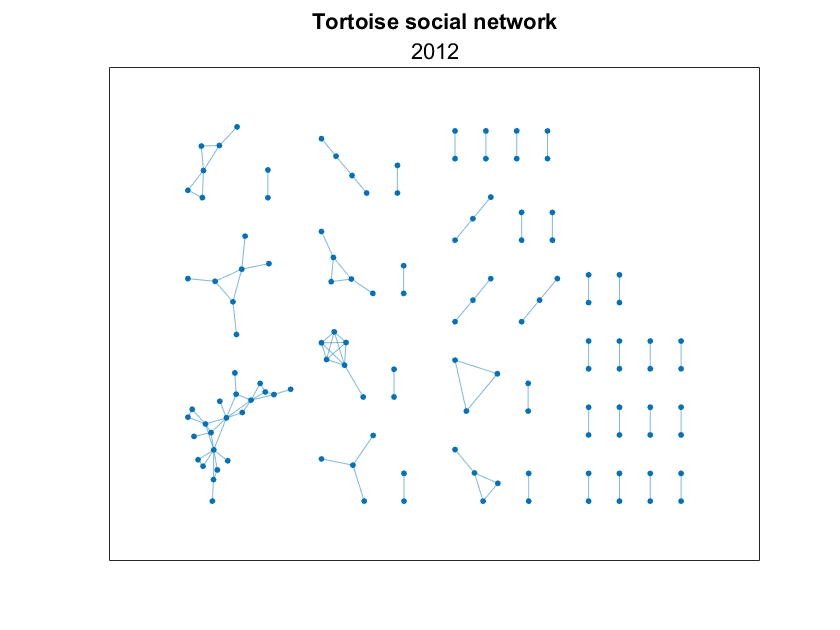}
     \includegraphics[width=0.23\textwidth]{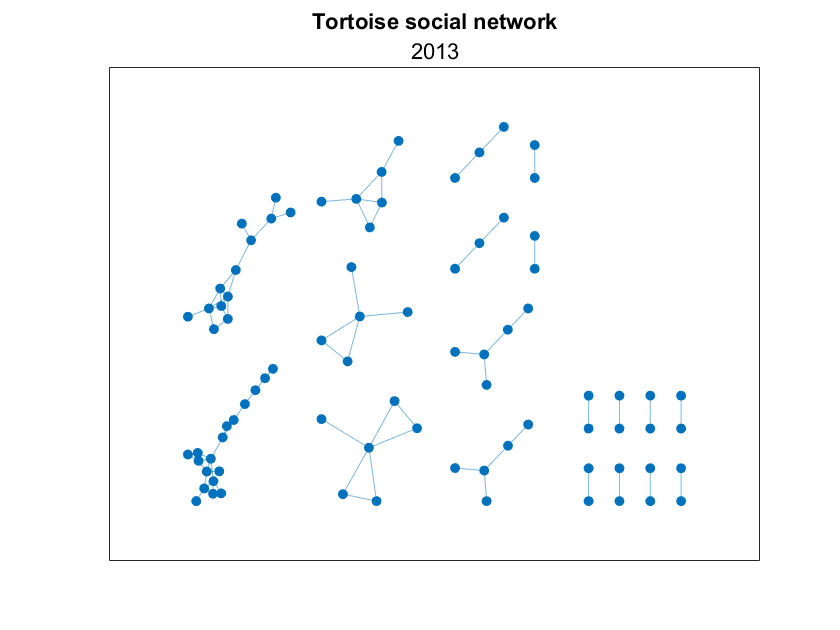}
  \end{center}
	\caption{\label{figure5} Social networks of desert tortoises derived from nest visits in some of the years from 2005 to 2012 at site FI (Fort Irwin, Mojave desert, from \cite{tortoise}). Data from the network repository \cite{data_repository}. }

\end{figure}

A precise mathematical approach to construct such association networks is proposed in \cite{tortoise} to study the use of refuges in a species of desert tortoises, Gopherus agassizii, by observational data of their asynchronous use of refuges. This species is highly solitary and the goal of the authors is mainly to understand the role of various factors, such as age, sex, and various stressors, on the preference for specific refuges. Here, however, we focus on their construction of the social network. The authors in \cite{tortoise} first construct a bipartite graph with two sets of nodes: the tortoises and the refuges, where each tortoise is associated to the set of refuges visited in a given year. Then, the social network is constructed by defining two tortoises as being mutually associated if they have visited some common refuge in a common time range. Formally, the social network is called the one-mode projection of the bipartite graph constructed above \cite{zweig}. The projected graphs have a peculiar structure, in that rare but relatively large cliques (subgraphs in which all agents are mutually connected) may occur (cf. Figure \ref{figure5}), there are no recognizable hubs and the degree distribution is not Poisson. In particular, the distribution of the cliques of the graph is most interesting, because these correspond to tightly knit groups of individuals, a feature uncharacteristic of many other social networks.

On the other hand, by construction, the one-mode projection of a bipartite graph is a simple graph, where the relations between nodes are binary, but  in association networks different types of associations may arise, since agents sharing the same resource are all in a joint relation to each other. For instance, it may be useful to distinguish the case in which, say, three agents are linked because each pair of them shares a common resource, different for each pair, from the case in which three agents share exactly the same resource. However, in both cases, the same 3-clique in the one-mode projection is induced, but it seems interesting to investigate the different instances.

This richer structure can be described in terms of hypergraphs, in which the nodes are linked by $n$-ary relations (cf., for instance, \cite{bretto} and, in the context of animal social networks, \cite{silk}). This is the natural language to study association networks since, under mild hypotheses, there is one-to-one correspondence between hypergraphs and graphs. Actually, an approach based on hypergraphs is more appropriate to study the spread of infections in association networks, relative to the conventional one based on projected graphs \cite{infection}.

A second important factor to be taken into account in the study of social networks is the intrinsic randomness of real networks. In the context of association networks, this can be done by using a class of random bipartite graphs and their one-mode projections introduced in  \cite{singer_cohen}, called 'random intersection graphs'. In the original approach of \cite{singer_cohen}, each agent independently chooses a random subset of a fixed set of resources, and two agents are related if they share at least one common resource. The structure of random intersection graphs has been extensively studied, for instance regarding the clique structure (cf. \cite{singer_cohen}, and the  generalizations \cite{generalized_RIG}, \cite{generalized_RIG_deg}, \cite{bloznelis1},  \cite{bloznelis4},\cite{communities}, \cite{epidemics}).  In Section 2 we review those notions of graph and hypergraph theory, as  well as of random intersection graphs, that are useful to our analysis. 

In our work we study the mathematical structure of social networks inspired by those introduced in \cite{tortoise}.  Our approach is as follows: fix a subset of points in the plane lattice, that play the role of resources or refuges, and consider a family of independent random walks, identified here to the agents: we construct a random bipartite graph as in \cite{tortoise}, by associating agents to sites visited in a finite, fixed amount of time.  From the mathematical point of view, this provides an instance of random intersection graph in which the random sets are subsets of the resources/sites, but where visits to different sites are not independent of each other. This yields a probability distribution on the sets of sites quite different from most variants of random intersection graphs, in which the probability distribution of the random sets is relatively simple.  In Section 3 we focus on some of the  features of random graphs associated to random walks.

Further, we have here the advantage that, for the simple random walk in the plane, most of the useful quantities can be computed explicitly or by iterative formulas, and this allows to fully characterize the random bipartite graphs and hypergraphs resulting from this procedure, and study their structure in dependence of the geometry of the point set representing the resources.  Our choice of working with simple random walks can be justified by studies of animal mobility that show that many species adopt search strategies for resources that resemble Brownian motion or L\'{e}vy walks \cite{pasquaretta}; indeed, a similar approach has been used in simulations of the foraging behavior of the spider monkey \cite{foraging}.  

We mention here that the study of intersections of random walks, initiated by Erd\H{os} and Taylor \cite{erdos_taylor}, has been developed mainly in the context of large-time asymptotics (cf. \cite{lawler_limic},  \cite{chen},\cite{baron}, \cite{ferraro}, \cite{common_sites}, \cite{visited_sites}): finite-time results are difficult due to the fact that intersection problems are intrinsically non-Markovian.  

In Section 4 we prove a set of useful results on 2-dimensional random walk that are relevant to our case. Finally, in Section 5, we apply the results of Section 4 to intersection networks of random walks, while Section 6 is devoted to a discussion of our model and to the conclusions.

As a final remark, we recall that social networks in which the association among the agents is determined by the common use of shared resources are widely used in a myriad of applications.  Examples range from cryptographic key-sharing protocols in server networks to affiliation networks (cf., e.g.,  \cite{breiger} for some of the first formulations of these models), such as coauthor networks for scientific publications (authors/scientific journals) or actor network (actors/movies).

\section{Preliminaries}
\label{preliminaries}

We briefly describe here the basic tools we use in this work, i.e.,  graphs, hypergraphs, and random intersection graphs. 

A  \textit{simple graph} is a pair $G=(V,E)$, with $V$ the set of nodes, or vertices, and $E\subseteq V\times V$ the set of edges. We assume that $V$ is finite with cardinality $N$ and label the nodes by Latin indices $\{1,\dots,i,\dots,N\}$ . We assume that $G$ is undirected, so that, for $i,j\in V$, an edge $(i,j)\in E$   if and only if $(j,i)\in E$, and we identify $(i,j)$ and $(j,i)$, in which case we say that $i$ and $j$ are adjacent. Also, we do not allow for self-loops, i.e., edges of the form $(i,i)$ for $i\in V$. The degree of node $i\in V$ is $\text{deg}_G(i)=|\{j\in V :(i,j)\in E\}|$, and the neighborhood of a node $i\in V$ is the set $\mathcal{N}(i)=\{j\in V : 
(i,j)\in E\}$, so that $\text{deg}_G(i)=|\mathcal{N}(i)|$.  The adjacency matrix of $G$ is the $N\times N$ symmetric matrix defined by 
\begin{equation}
 A_{ij}=\begin{cases}
     1 & \text{if } (i,j)\in E
     \\
     0 & \text{otherwise},
 \end{cases}   
\end{equation}
for $i,j=1,\dots,N$.
A simple graph is {\it complete} if $E=V\times V$, i.e., if all nodes are adjacent. A {\it clique} in a graph $G$ is a complete induced subgraph, i.e., is a set of vertices that are all adjacent in $G$. A maximal clique is a clique that is not contained in a larger clique.

A \textit{bipartite graph} is a graph $B=(V_1\cup V_2, E_B)$ in which the set of vertices is the union of two  sets $V=V_1\cup V_2$, with $V_1\cap V_2=\emptyset$, such that each node in $V_1$ is only adjacent to nodes in $V_2$ and conversely, and there are no edges between nodes in $V_1$,  and no edges between nodes in $V_2$ as well. If we label the nodes so that $V_1=\{1,\dots,N\}$ and $V_2=\{N+1,\dots,N+M\}$ (so that $V$ has now cardinality $N+M$) and use Latin and Greek indices for nodes in $V_1$ and $V_2$, respectively, the adjacency matrix of a bipartite graph has the form 
\begin{equation}
    A=\left(\begin{array}{cc}
       0  &  J\\
       J^\top  & 0
    \end{array}\right),
\end{equation}
with $J$ the so-called biadjacency matrix, a $N\times M$ matrix defined as $J_{i\alpha}=1$ when $(i,\alpha)\in E_B$, with $i\in V_1$ and $\alpha\in V_2$, and $\top$ is the transpose. 

A \textit{hypergraph} is a pair $H=(V,F)$, with $V$ the set of nodes, labeled as before by Latin indices  $i\in\{1,\dots,N\}$, and $F$, the set of  hyperedges,  is a collection of subsets of $V$, possibly empty or singletons. Two vertices in $V$ are adjacent if there is a hyperedge containing both, while we say that two hyperedges are incident if their intersection is not empty.  If we denote the hyperedges by $F=\{f_1,\dots,f_\alpha,\dots,f_M\}$, the incidence matrix of $H$ is the $N\times M$ matrix $ \tilde J$ such that, for $i=1,\dots,N$ and $\alpha=1,\dots,M$,
\begin{equation}
 \tilde J_{i\alpha}=\begin{cases}
     1 & \text{if } i\in f_\alpha
     \\
     0 & \text{otherwise}.
 \end{cases}   
\end{equation}
Also, the adjacency matrix  of a hypergraph is the $N\times N$ symmetric matrix $\tilde A$ such that
\begin{equation}
 \tilde A_{ij}=\begin{cases}
     \sum_{\alpha=1}^M\tilde J_{i\alpha}\tilde J_{j\alpha} & \text{if } i\ne j
     \\
     0 & \text{otherwise},
 \end{cases}   
\end{equation}
for $i,j=1,\dots,N$. Notice that  $\tilde A_{ii}=0$. Hence, a non-diagonal element of $\tilde A$ is the number of mutually incident hyperedges containing two distinct nodes $i,j$. The adjacency matrix of a hypergraph can be viewed as the adjacency matrix of a multigraph with nodes $V$, where a multigraph is a graph with possibly repeated edges.

To each hypergraph is associated a bipartite graph, called the incidence graph. The bipartite graph has vertex set $V\cup F$, and the edges are pairs $(i,\alpha)$, with $i\in f_\alpha$ for $i\in V$. In this representation, every hyperedge corresponds to a vertex in the set $F$, and a vertex $i$ in $V$ is adjacent to a vertex in $F$ if and only if $i$ is contained in the corresponding hyperedge. 

Conversely, to each bipartite graph  $B=(V\cup F, E_B)$ with $V=\{1,\dots,i,\dots,N\}$, $F=\{1,\dots,\alpha,\dots,M\}$ and edge set $E_B$, a unique hypergraph is associated, where the hyperedges $f_\alpha$ for $\alpha\in F$ are defined by $f_\alpha=\{i\in V : (i,\alpha)\in E_B\}$.    In other words, each vertex $\alpha\in F$  corresponds to a hyperedge $f_\alpha$ that contains all nodes in $V$ that are adjacent to $\alpha$. 

Notice that, in this formulation, the incidence matrix $\tilde J $ of the hypergraph coincides with the biadjacency matrix $J$ of the bipartite graph, and conversely. Also, we allow here the hypergraph to have repeated hyperedges, i.e., there may exist $\alpha,\alpha'$ such that $f_\alpha=f_{\alpha'}$.

The \textit{one-mode projection}  onto $V$ of a bipartite graph $B=(V\cup F,E_B)$ is the simple graph $(V, E)$ such that two nodes $i,j\in V$ are adjacent in $E$ if and only if, when viewed as nodes of the bipartite graph, they are adjacent to at least a common node in $F$, i.e., $(i,j)\in E$ if and only if there exists $\alpha\in F$ such that $(i,\alpha),(j,\alpha)\in E_B$. Hence, the adjacency matrix of the one-mode projection is  
\begin{equation}
 A_{ij}=\begin{cases}
   1   & \text{if }\sum_{\alpha=1}^MJ_{i\alpha}J_{j\alpha}>0
     \\
     0 & \text{otherwise},
 \end{cases}   
\end{equation}
for $i,j=1,\dots,N$.  The same procedure can be used to define the one-mode projection on $F$. Notice that the nodes in $V$ that are adjacent to the same node $\alpha \in F$ form a clique in the one-mode projection $G$ on $V$.

Given a hypergraph $H=(V,F)$, its \textit{2-section} is the simple graph $G=(V,E)$ with the same vertex set, and where two vertices are adjacent in $G$ if and only if they are adjacent in $H$, that is, they belong to the same hyperedge. The adjacency matrix of $G$ is defined as
\begin{equation}
 A_{ij}=\begin{cases}
   1   & \text{if } \exists f_\alpha \in F \text{ such that } i,j\in f_\alpha
     \\
     0 & \text{otherwise},
 \end{cases}   
\end{equation}
for $i,j=1,\dots,N$.  The vertices belonging to the same hyperedge are mutually connected and therefore form a clique in the 2-section $G$. 

 Notice that the one-mode projection of a bipartite graph coincides with the 2-section of the associated hypergraph. 
\begin{figure}[t]
	\begin{center}
 \includegraphics[width=.9\textwidth]{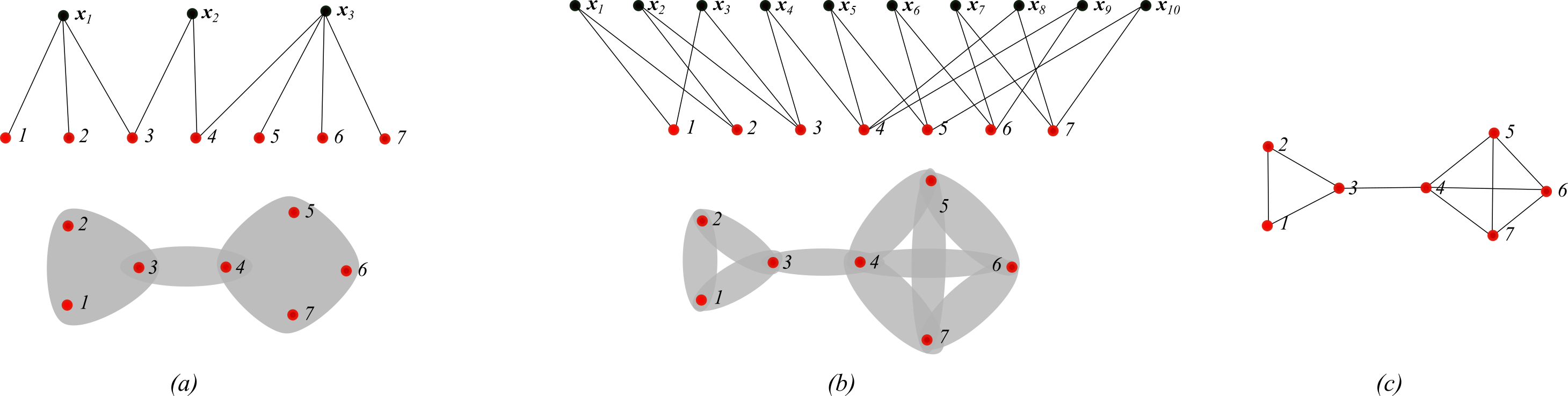}
	\end{center}
	\caption{\label{figure3} (a) A bipartite  graph on 3 sites (top) and the corresponding  hypergraph (bottom);  (b) a bipartite  graph on 7 sites (top) and the corresponding  hypergraph (bottom); (c) the one-mode projections (2-section) of the graphs in (a) and (b) are the same. }

\end{figure}

To summarize, in the above paragraphs we have introduced four classes of objects: hypergraphs and bipartite graphs, which we regard here as equivalent constructs, and the corresponding 2-sections and one-mode projections, respectively, which we also regard as equivalent. Two simple examples are sketched in Figure \ref{figure3}. Notice that different bipartite graphs (or hypergraphs) may have the same one-mode projection (2-section), so that projections may lose information with respect to the original structure.

A \textit{random graph} (or \textit{random hypergraph}) $G$ is a graph (hypergraph) randomly selected from a set of graphs $\mathcal G$, according to an assigned probability measure $\pp$ such that, for each $G\in\mathcal G$, $\pp(G)$ is the probability that the specific graph $G$ is realized. Graph functions such as degree, clique size, etc., become random variables with their own distribution functions.

A special class of random graphs are random intersection graphs, defined as follows \cite{singer_cohen}.  Let $V$ and $F$ be finite sets of cardinality $N$ and $M$, respectively, and,  to each $i\in V$ assign a random set 
\begin{equation}   
e_i=\{\alpha_1,\dots\alpha_k\}\subset F,
\end{equation}  
independently on $i$. The random intersection graph  is the simple graph $G=(V,E)$, with nodes $V$ and edges defined by 
\begin{equation}
    (i,j)\in E \text{ if and only if } e_i\cap e_j\ne \emptyset,
\end{equation}
for $i,j=1,\dots,N$.  

This construction defines indeed a random bipartite graph $B=(V\cup F,E_B)$, with edges
\begin{equation}
    (i,\alpha)\in  E_B \text{ if and only if } \alpha\in e_i,
\end{equation}
for $i=1,\dots, N$ and  $\alpha=1,\dots, M$ . In other words, each of the objects  $\alpha\in F$ is viewed as a vertex in $F$, and is adjacent to all nodes in $V$ that have 'chosen' it. Notice that $G$ is the one-mode projection of  $B$ on $V$. 

The above procedure equivalently defines a random hypergraph  $H=(V,F)$ where, with abuse of notation, we denote by $F=(f_1,\dots,f_\alpha,\dots,f_M)$  the set of hyperedges, given by
\begin{equation}
    f_\alpha=\{i\in V : \alpha\in e_i\},  
\end{equation}
$\alpha=1,\dots, M$. In other words, the hyperedge $f_\alpha$ is the set of nodes $i\in V$ that have 'chosen' $\alpha$. Notice that $G$ is the 2-section of $H$.

Since, by assumption, in a random intersection graphs we assume that the assignment of the random sets $e_i$ are independent on $i\in V$, the probability of the associated random bipartite graph $B$, as well as of its corresponding hypergraph $H$, is simply
\begin{equation}
\pp(B)=\pp(H)=\prod_{i=1}^N\pp(e_i).
\label{prob_bipartite_1}
\end{equation}

\section{The intersection network of random walks} 

Our goal is to construct a social network in which the nodes correspond to agents moving at random on the plane square lattice, and such that two agents are connected if they have visited the same sites in a finite time interval, where the sites belong to a given assigned set. For instance, in the situation studied in \cite{tortoise}, the agents are the tortoises and the sites the refuges. 

A possibility to define the intersection network of a set of random walks is to work directly in terms of  intersection graphs \cite{tortoise}. However, for many applications, it is better to exploit the rich structure of the underlying hypergraph, or the corresponding bipartite graph, and derive the intersection network from them. We reintroduce these structures in the next subsections, referring specifically to the special case of the intersection network of a set of random walks and specifying the probabilities of interest for these graphs.

We denote by $S=(S_t)_{t=0,1,\dots}$ the time-discrete simple random walk on $\zz^2$, with starting point the origin, and uniform transition probability: for each $\bfx=(h,k)$ and $\bfy=(h',k')$, it holds
\begin{equation*}
\pp(S_{t+1}=\bfy|S_{t}=\bfx)=
\begin{cases}
1/4  &\text{if  }h'=h\pm1\text{ or } k'=k\pm1
\\
0&\text{otherwise}.
\end{cases}
\end{equation*}
We write
\begin{equation}
S[1,T]=\{S_t:1\le t\le T\}
\label{range}
\end{equation}
for the set of points visited by the random walk up to time $T$, excluding the initial point. 

We denote by 
\begin{equation*}
S^1,\dots,S^N,
\end{equation*} 
a set of $N$ independent simple random walks on $\zz^2$ starting at the origin.  We  use Latin letters $i,j,\dots\in\{1,\dots,N\}$ to label each random walk.  We shall often refer to the random walks as the agents, and denote by $V=\{1,\dots,N\}$ the set of agents.

Further, we recall the definition of the first arrival time at $\bfx\in \zz^2$ of a random walk starting at $\bfx_0 \in\zz^2$ 
\begin{equation*}
    \tau^{\bfx_0}_{\bfx}=\min\{t\in \nn : S_t=\bfx| S_0=\bfx_0\}\qquad\text{and}\qquad \tau_{\bfx}=
    \tau_{\bfx}^{\bfzero}.
\end{equation*}
When needed, and the starting point is the origin, and it is necessary to indicate which random walk we refer to, we shall drop the superscript denoting the starting point,  and write, with a slight abuse of notation,
\begin{equation*}
    \tau^{(i)}_{\bfx}=\min\{t\in \nn : S_t^i=\bfx| S^i_0=\bfzero\}.
\end{equation*}
Let
\begin{equation}
    F=\{\bfx_1,\dots,\bfx_\alpha,\dots,\bfx_M\}\subset \zz^2
    \label{sites}
\end{equation}
be a finite and fixed set of points in the square lattice, which represents the set of relevant sites whose common visits by the random walks determine the association among the agents. For $A=\{\alpha_1,\dots,\alpha_k\}\subseteq\{1,\dots,M\}$ and $k=1,2,\dots, M$ a set of indices, we write 
\begin{equation*}
\bfx_{A}=\{\bfx_{\alpha_1},\dots,\bfx_{\alpha_k}\}\subseteq F,\qquad
\bfx_{-A}=
\{\bfx_{\beta_1},\dots,\bfx_{\beta_{M-k}}\},
\end{equation*}
such that $\bfx_{A}\cap\bfx_{-A}=\emptyset$ and $\bfx_{A}\cup\bfx_{-A}=F$.

\subsection{The intersection bipartite graph}

We now define a random bipartite graph whose structure is determined by the interactions of agents visiting a fixed set $F$ of sites. 
We consider a random bipartite graph whose vertex set $V\cup F$ is such that \( V \) is the set of agents and \( F \) is the set of sites, and connect an agent to a site if and only if the agent has visited the site within the interval \( [1,T] \). We model the agents' displacement as independent two-dimensional random walks, so that the randomness of such a bipartite graph is determined by the random motion of the agent.

Consider first a single random walk and define a random set on $F$ as the set formed by the sites in $F$ visited by the random walk in the time interval $[1,T]$: 
\begin{equation}
 e:= \{\bfx_\alpha\in F:\bfx_\alpha\in S[1,T]\}=S[1,T]\cap F.
\end{equation}
 By definition, the probability that a given set $e=\{\bfx_{\alpha_1},\dots,\bfx_{\alpha_k}\}=\bfx_{A}\subset F$ be realized is given by
\begin{equation}
\pp(e)
=\pp(\bfx_A\subset S[1,T], \bfx_{-A}\cap S[1,T]=\emptyset),
\label{Q_definition_1}
\end{equation}
which is the probability that the random walk, starting at the origin, visits the sites in the set $\bfx_{A}$ in the interval $[1,T]$, avoiding the remaining sites in $\bfx_{-A}$.  
A related notion is the probability that random  walk visits the sites in  $\bfx_{A}$ in the interval $[1,T]$, possibly visiting also the remaining sites in $F$: for $\bfx_A\subseteq F$, we write
\begin{equation}
  Q_{\bfx_A}(T):=\pp(\bfx_{A}\subset S[1,T]).
   \label{marginals}
\end{equation}
Notice that from \eqref{marginals} it follows that
\begin{equation}
\bfx_A \subseteq  \bfx_{A'}\quad \Rightarrow\quad  Q_{\bfx_{A'}}(T)\le
Q_{\bfx_{A}}(T).
   \label{inclusion}
\end{equation}
We show below in Proposition \ref{prop_marginals} that the probability distribution in \eqref{Q_definition_1} is completely determined by \eqref{marginals}, but elementary arguments allow to discuss a simple case. 
\begin{example}
\label{example1}
For $F=\{\bfx_1,\bfx_2,\bfx_3\}$ and $A=\{2\}$ and $-A=\{1,3\}$, we have
\begin{equation*}
  \pp(\{\bfx_2\})=  Q_{\{\bfx_2\}}(T)-Q_{\{\bfx_1,\bfx_2\}}(T)-Q_{\{\bfx_2,\bfx_3\}}(T)
  +Q_{\{\bfx_1,\bfx_2,\bfx_3\}}(T),
\end{equation*}
where $Q_{\{\bfx_2\}}(T)=\pp(\bfx_2\in S[1,T])$. The above result can be proved by noting that 
\begin{equation*}
\begin{split}
\pp(\{\bfx_2\})&= \pp(\bfx_2\in S[1,T],\bfx_1,\bfx_3\notin S[1,T])   
  \\
  &=   \pp(\bfx_2\in S[1,T],\bfx_3\notin S[1,T])-
  \pp(\bfx_1,\bfx_2\in S[1,T],\bfx_3\notin S[1,T])
  \\
  &= \pp(\bfx_2\in S[1,T])-
  \pp(\bfx_1,\bfx_2\in S[1,T])-\pp(\bfx_2,\bfx_3\in S[1,T])
  + \pp(\bfx_1,\bfx_2,\bfx_3\in S[1,T]).
  \end{split}
\end{equation*}

\end{example}

Consider now a set of agents $V$, i.e., a list $\{S^1,\dots,S^N\}$  of $N$ independent simple random walks on $\zz^2$ starting at the origin. For each $i=1,\dots,N$, independently in $i$, define the random set $e_i$ as the set of sites visited by $S^i$ up to time $T$: 
\begin{equation}
e_i := \{\bfx_\alpha\in F:\bfx_\alpha\in S^i[1,T]\}  ,\qquad i=1,\dots,N,
\label{e_i_definition}
\end{equation}
 where 
\begin{equation}
S^i[1,T]=\{S^i_t:1\le t\le T\},\qquad i=1,\dots,N.
\label{range_multiple}
\end{equation}
We can now define the random bipartite graph $B=(V\cup F,E_B)$, with edges defined by
\begin{equation}
(i,\alpha)\in E_B \quad \text{ if and only if }\quad
\bfx_\alpha\in e_i,
\end{equation}
for $i\in V=\{1,\dots,N\}$ and $\alpha\in \{1,\dots,M\}$. 
Notice that, in this formulation, no account is taken of the number of times that a random walk visits the same site.  
For $\bfx_\alpha\in F$, we also define the random set 
\begin{equation}
f_\alpha=\{i\in V :\bfx_\alpha\in e_i\},
\end{equation}
which is the set of agents that have visited site $\alpha$.  Notice that, for $i\in V$, $\deg_B(i)=|e_i|$ is the number of sites visited by the random walk, while for $\bfx_\alpha\in F$, $\deg_B(\alpha)=|f_\alpha|$ is the number of agents that have visited $\bfx_\alpha$. 

\begin{figure}[ht]
	\begin{center}
    \includegraphics[width=.9\textwidth]{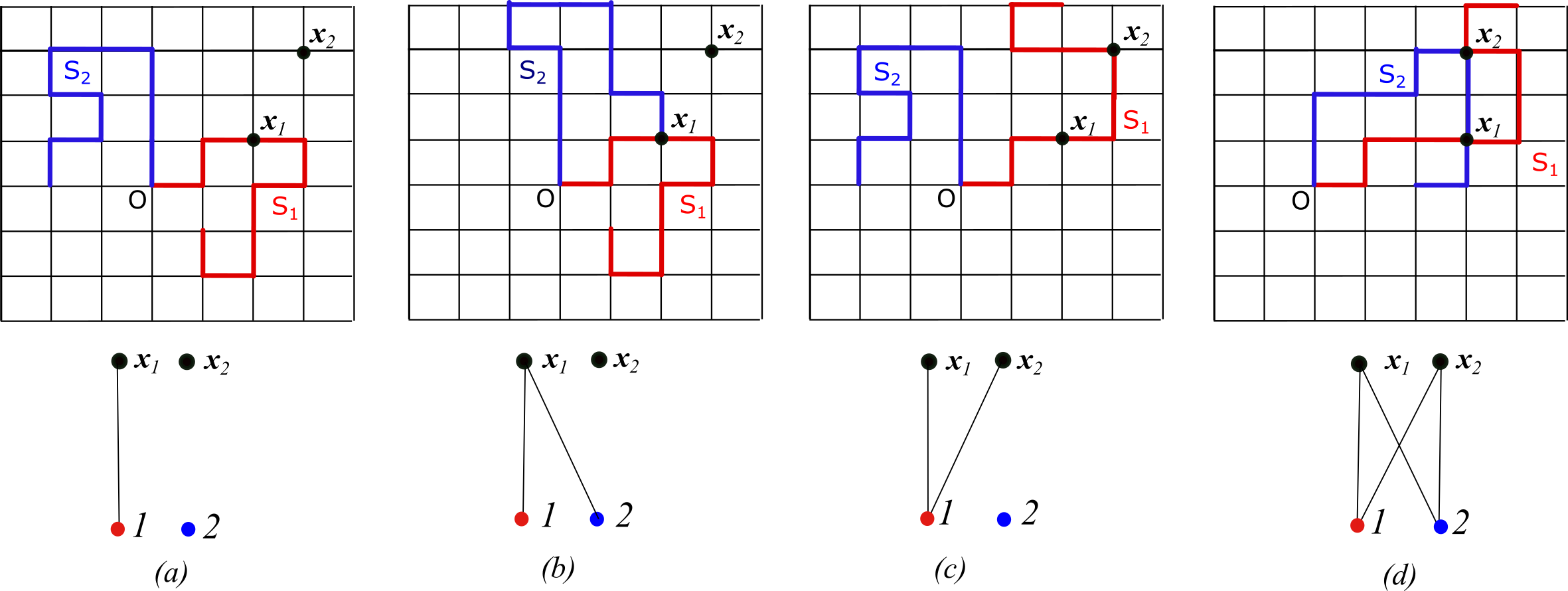}
	\end{center}
	\caption{\label{figure2} The bipartite graphs in Example \ref{example2} and the corresponding random walks: (a) $B_1$, (b) $B_2$,
(c) $B_3$, (d) $B_4$.  }
\end{figure}

Since a random bipartite graph $B$ is completely defined by the random sets $e_i$, and the random walks are independent, its  probability distribution is given by \eqref{prob_bipartite_1}:
\begin{equation}
\pp(B)=\prod_{i=1}^N\pp(e_i).
\label{full_Q}
\end{equation}

Hence, as a consequence of independence, the probability distribution of the intersection bipartite graph as above is completely determined by the distribution \eqref{Q_definition_1} of a single random walk. In the next section, we introduce analytical results for its evaluation, along with methods to compute other interesting quantities that characterize the considered graphs.

\begin{example}\label{example2}
Let $F=\{\bfx_1,\bfx_2\}$ and $V=\{1,2\}$. In this simple case we can compute the probabilities that each of the 4 bipartite graphs $B_1,\dots,B_4$ in Figure \ref{figure2} are realized, where the edges are
\begin{equation*}
 E_{B_1}=\{(1,1)\},
 E_{B_2}=\{(1,1),(2,1)\},
 E_{B_3}=\{(1,1),(1,2)\}
, E_{B_4}=\{(1,1),(1,2),(2,1),(2,2).
\end{equation*}
We have 
\begin{equation*}
\begin{split}
\pp(B_1)  &= (Q_{\{\bfx_1\}}(T)- Q_{\{\bfx_{1},\bfx_2\}}(T))
(1+ Q_{\{\bfx_{1},\bfx_2\}}(T))
-Q_{\{\bfx_1\}}(T)- Q_{\{\bfx_2\}}(T))
\\
\pp(B_2)  &= (Q_{\{\bfx_1\}}(T)- Q_{\{\bfx_{1},\bfx_2\}}(T))^2
\\
\pp(B_3)  &= Q_{\{\bfx_{1},\bfx_2\}}(T)
(1+ Q_{\{\bfx_{1},\bfx_2\}}(T)
-Q_{\{\bfx_1\}}(T)- Q_{\{\bfx_2\}}(T))
\\
\pp(B_4)  &= 
(Q_{\{\bfx_{1},\bfx_2\}}(T))^2.
\end{split}
\end{equation*}
To see this, consider first $B_1$: by \eqref{full_Q},  
$\pp(B_1)=\pp(e_1) \pp(e_2)$ and, in this case,  $e_1=\{\bfx_1\}$ and $e_2=\emptyset$. On the other hand, either by Proposition \ref{prop_marginals} or by a direct computation as in Example \ref{example1},
\begin{equation*}
\pp(\{\bfx_1\})=Q_{\{\bfx_1\}}(T)- Q_{\{\bfx_{1},\bfx_2\}}(T),
\end{equation*}
and 
\begin{equation*}
\pp(\emptyset)= 1+ Q_{\{\bfx_{1},\bfx_2\}}(T)
-Q_{\{\bfx_1\}}(T)- Q_{\{\bfx_2\}}(T).
\end{equation*}
As to the remaining graphs, it is enough to notice that, for  $B_2$,
$e_1=e_2=\{\bfx_1\}$, while for $B_3$,  $e_1=\{\bfx_1,\bfx_2\}$ and $e_2=\emptyset$, and finally, for $B_4$, $e_1=e_2=\{\bfx_1,\bfx_2\}$.

Notice that, for $B_1$, we have $f_1=\{1\}$ and $f_2=\emptyset$; for $B_2$, $f_1=\{1,2\}$ and $f_2=\emptyset$;  for $B_3$, $f_1=\{1\}$ and $f_2=\{1\}$; and for $B_4$, $f_1=\{1,2\}$ and $f_2=\{1,2\}$.
\end{example}



\subsection{The intersection hypergraph}

While the bipartite graph $B$ is defined in terms of the sites visited by each agent, reformulating the model in term of the associated hypergraph allows to focus on the relations among the agents that have visited the same sites.  

Given as before a set of agents $V$ and a set of sites $F$, and recalling Section  \ref{preliminaries}, we define the family of random sets on $V$ as
\begin{equation}
f_\alpha:=
\{i\in V:\bfx_\alpha\in S^i[1,T]\} \subset V 
\end{equation}
for each $\alpha\in\{1,\dots,M\}$ or, equivalently, 
\begin{equation}
f_\alpha:=
\{i\in V:\bfx_\alpha\in e_i\}\subset V, 
\end{equation}
where $e_i$ is defined in \eqref{e_i_definition}. In terms of the situation described in \cite{tortoise}, $f_\alpha$ and  $|f_\alpha|$ are the set and the number of tortoises that have visited the refuge labelled by $\alpha$.

Since the random walks are independent, the probability of a random set $f_\alpha=\{i_{1},\dots,i_{K}\}$ of cardinality $K$ is
\begin{equation}
\pp(f_\alpha)=Q_{\bfx_\alpha}(T)^K(1-Q_{\bfx_\alpha}(T))^{N-K}.
\label{probability_alpha_face}
\end{equation}
To see this, notice that
\begin{equation*}
\begin{split}
\pp(f_\alpha)&=
\pp(\bfx_\alpha \in S^{i_1}[1,T],\dots, \bfx_\alpha \in S^{i_K}[1,T], 
\bfx_\alpha \notin S^{i_{K+1}}[1,T],\dots, \bfx_\alpha \notin S^{i_{N}}[1,T])
\\&
=\Pi_{i={i_1}}^{i_K}\pp(\bfx_\alpha \in S^i[1,T])
\Pi_{j={i_{K+1}}}^{i_N}\pp(\bfx_\alpha \notin S^j[1,T])\\&
=\pp(\bfx_\alpha \in S[1,T])^K(1-\pp(\bfx_\alpha \in S[1,T])^{N-K},
\end{split}
\end{equation*}
where we have used \eqref{marginals},  and where $\{i_{K+1},\dots,i_{N}\}$ are the indices of agents in the complement of $f_\alpha$ in $V$.

The intersection hypergraph is just the random hypergraph $H=(V,F)$ with hyperedges $(f_\alpha)_{\{1,\dots,M\}}$. We remark that each hyperedge $f_\alpha$ is the set of agents that have visited site $\bfx_\alpha$ in the time interval $[1,T]$. To each site $\bfx_\alpha$ is therefore associated a unique hyperedge $f_\alpha$.

Let $\mathcal H$ be the set of hypergraphs with vertex set $V$ and $M$ hyperedges. The probability distribution of a random hypergraph in this set cannot be written simply in terms of the probability of the hyperedges, because the event that a random walk hits a given point is not independent of the event of hitting another fixed point. However, recalling that there is a unique bipartite graph $B$ associated to each hypergraph $H$, and conversely, we can write
\begin{equation}
\pp(H)=\pp(B)=\Pi_{i=1}^N\pp(e_i).
\label{full_H}
\end{equation}

\subsection{2-sections and one-mode projections}

We define here the intersection graph of the $N$ random walks as a simple graph on $V$ in which two agents are adjacent if and only if they have visited the same sites in the time interval $[1,T]$.
As mentioned in Section \ref{preliminaries}, this can be derived either as the one-mode projection on $V$ of the intersection bipartite graph $B$ or, equivalently, as the 2-section of the intersection hypergraph $H$. 
Formally, the intersection graph is the random simple graph $G=(V,E)$ such that, for $i,j\in V$,
\begin{equation}
(i,j) \in E \quad\text{ if and only if }\quad F\cap S^i[1,T]\cap S^j[1,T]\ne\emptyset.  
\end{equation}
Equivalently, viewing $G$ as the one-mode projection of $B$
\begin{equation}
(i,j) \in E \quad\text{ if and only if }e_i\cap e_j\ne\emptyset,  
\label{adjacency_agents}
\end{equation}
and, viewing instead $G$ as the 2-section of $H$,
\begin{equation}
(i,j) \in E \quad\text{ if and only if }\quad\exists \bfx_\alpha\in F \text{ such that }
i,j\in f_\alpha.
\end{equation}
Therefore, each site $\bfx_\alpha\in F$, or, equivalently, each node $\alpha$  of $F$ in $B$ or, again equivalently, each hyperedge $f_\alpha$ of $H$, induces a clique in $G$ of order 
\begin{equation*}
    \deg_B(\alpha)=|f_\alpha|.
\end{equation*}
However, the derivation of the probability distribution for the random graph $G$ from the corresponding distribution functions of $B$ and $H$ is not straightforward, since many different intersection bipartite graphs, or hypergraphs,  project to the same intersection graph on $V$, as shown in Figure \ref{figure3}.

\section{Analytical results}
\label{sec_analytical_results}

In this paragraph, we prove a set of results related to the computation of the probabilities of the one mode projections of the random graph 
$G$. As mentioned above, we cannot provide a general formula for these probabilities due to the different combinatorial challenges in describing specific instances. Therefore, we limit ourselves to determining the expressions for the involved probabilities.

Our first aim is to relate  the probability distribution of the random sets in \eqref{Q_definition_1} with the probability \eqref{marginals} that the random walk, starting at $\bfzero$,  visits all the sites in $\bfx_A$, possibly also visiting the remaining sites in $F$. We will show later that \eqref{marginals}, in turn, can be computed by iterative formulas involving various generalizations of the distribution of first arrival times (cf.  Theorem \ref{general_rep_Q} in Section \ref{subsec_analytical results}).

\begin{proposition}
\label{prop_marginals}
The probability distribution in \eqref{Q_definition_1} is uniquely determined by the functions in \eqref{marginals}, i.e., for $\bfx_A\subseteq F$, writing $A=\{\alpha_1,\dots,  \alpha_k \}$ and $-A=\{1,\dots,M\}\setminus A=\{ \beta_1,\dots,  \beta_{M-k} \}$, we have 
\begin{equation}
   \pp(\bfx_A)=\sum_{n=0}^{M-k}(-1)^n\sum_{\beta_{i_1}<\dots<\beta_{i_n}}
   Q_{\bfx_{A\cup\{\beta_{i_1},\dots,\beta_{i_n}\}}}(T) ,
   \label{marginals1}
\end{equation}
where the sum at the right-hand side is taken on all ordered $n$-tuples of indices in $-A$. In particular, 
\begin{equation}
   \pp(F)=
   Q_{\{\bfx_1,\dots,\bfx_M\}}(T)= Q_{F}(T),
   \label{marginals2}
\end{equation}
and
\begin{equation}
   \pp(\emptyset)= 1+ \sum_{n=1}^{M}(-1)^n\sum_{\beta_{i_1}<\dots<\beta_{i_n}}
   Q_{\{ \bfx_{\beta_{i_1}},\dots,\bfx_{\beta_{i_n}}\}}(T).
   \label{marginals3}
\end{equation}
\end{proposition}

\begin{corollary}
    
    Notice that \eqref{marginals1} can be written in the more compact form
\begin{equation}
   \pp(\bfx_A)=\sum_{\bfx_A'\supseteq\bfx_A} (-1)^{|\bfx_A'|-|\bfx_A|}
   Q_{\bfx_{A'}}(T) ,
   \label{marginals1_1}
\end{equation}
where     $|\bfx_A|$ is the cardinality of $\bfx_A$.
\end{corollary}

\begin{proof} The assertion is akin to the inclusion/exclusion formula, but we prove it here explicitly, by induction on the number of points not visited by the random walk. Writing  (notice the change of notation)
$A=\{\alpha_1,\dots,  \alpha_{M-h} \}$ and $-A=\{ \beta_1,\dots,  \beta_{h} \}$, we use induction on $h$, i.e. the cardinality of the complement of $\bfx_A$ in $F$. To keep the notation simple, throughout the proof we write
$$
f( \bfx_{\alpha_{1}},\dots,\bfx_{\alpha_{k}}):=
Q_{\{ \bfx_{\alpha_{1}},\dots,\bfx_{\alpha_{k}}\}}(T).
$$
Now, for $h=1$ and for every $M> 1$, the thesis is true, since for every choice of $M$ and indices
$\{\alpha_1,\dots,  \alpha_{M-h},\beta_1 \}$
\begin{equation*}
\begin{split}
    \pp(\bfx_A)=&\pp(\bfx_{\alpha_1},\dots,\bfx_{\alpha_{M-1}}\in S[1,T],\bfx_{\beta_1}\notin S[1,T])
   \\
   =& \pp(\bfx_{\alpha_1},\dots,\bfx_{\alpha_{M-1}}\in S[1,T])
   - \pp(\bfx_{\alpha_1},\dots,\bfx_{\alpha_{M-1}},\bfx_{\beta_1}\in S[1,T])
   \\
   =&f(\bfx_{\alpha_1},\dots,\bfx_{\alpha_{M-1}})-
f(\bfx_{\alpha_1},\dots,\bfx_{\alpha_{M-1}},\bfx_{\beta_1}).
   \end{split}
\end{equation*}
We now prove that if the assertion is true for $h-1$ and every $M$, then it is true for 
$h$ and every $M$. Assume that $h>1$: then 
\begin{equation*}
\begin{split}
   \pp(\bfx_{\alpha_1},\dots,\bfx_{\alpha_{M-h}}\in S[1,T],&\bfx_{\beta_1},\dots,\bfx_{\beta_{h}}\notin S[1,T])
   \\
   =&
   \pp(\bfx_{\alpha_1},\dots,\bfx_{\alpha_{M-h}}\in S[1,T],\bfx_{\beta_1},\dots,\bfx_{\beta_{h-1}}\notin S[1,T])
   \\
   &- 
   \pp(\bfx_{\alpha_1},\dots,\bfx_{\alpha_{M-h}},\bfx_{\beta_{h}}\in S[1,T],\bfx_{\beta_1},\dots,\bfx_{\beta_{h-1}}\notin S[1,T]).
   \end{split}
\end{equation*}
By the inductive hypothesis, the assertion is true for $A=\{\bfx_{\alpha_1},\dots,\bfx_{\alpha_{M-h}}\}$ and $F=\{\bfx_{\alpha_1},\dots,\bfx_{\alpha_{M-h}},\bfx_{\beta_1},\dots,\bfx_{\beta_{h-1}}\}$, 
which has cardinality $M-1$; hence,
\begin{equation*}
\begin{split}
   \pp(\bfx_{\alpha_1},\dots,\bfx_{\alpha_{M-h}}&\in S[1,T],\bfx_{\beta_1},\dots,\bfx_{\beta_{h-1}}\notin S[1,T])
   \\
   =&
   \sum_{n=0}^{h-1}(-1)^n\sum_{\beta_{i_1}<\dots<\beta_{i_n}}
   f( 
   \bfx_{\alpha_1},\dots,\bfx_{\alpha_{M-h}},
\bfx_{\beta_{i_1}},\dots,\bfx_{\beta_{i_n}}),
   \end{split}
\end{equation*}
where the sum at the right-hand side is taken on all ordered subsets of $\{\bfx_{\beta_1},\dots,\bfx_{\beta_{h-1}}\}$, and can be viewed as the sum on all ordered subsets of 
$\{\bfx_{\beta_1},\dots,\bfx_{\beta_{h}}\}$, with cardinality $h-1$,  that do not contain $\bfx_{\beta_{h}}$. 
Further, 
\begin{equation*}
\begin{split}
   \pp(\bfx_{\alpha_1},\dots,\bfx_{\alpha_{M-h}},\bfx_{\beta_{h}}&\in S[1,T],\bfx_{\beta_1},\dots,\bfx_{\beta_{h-1}}\notin S[1,T])
   \\
   =&
\sum_{n=0}^{h-1}(-1)^n\sum_{\beta_{i_1}<\dots<\beta_{i_n}}
   f( 
   \bfx_{\alpha_1},\dots,\bfx_{\alpha_{M-h}},\bfx_{\beta_{h}},
\bfx_{\beta_{i_1}},\dots,\bfx_{\beta_{i_n}}) 
   \end{split}
\end{equation*}
where again the sum at the right-hand side of the above formula is taken on all ordered subsets of $\{\bfx_{\beta_1},\dots,\bfx_{\beta_{h-1}}\}$, and can be viewed as the sum on all ordered subsets of 
$\{\bfx_{\beta_1},\dots,\bfx_{\beta_{h}}\}$, with cardinality $h$, that contain $\bfx_{\beta_{h}}$, so that we can  write the right-hand side of the above formula as
\begin{equation*}
-\sum_{n=0}^{h}(-1)^n\sum_{\substack{\beta_{i_1}<\dots<\beta_{i_n}
\\
\beta_{h}\in\{\beta_{i_1},\dots,\beta_{i_n}\}
}}
   f( 
   \bfx_{\alpha_1},\dots,\bfx_{\alpha_{M-h}},
\bfx_{\beta_{i_1}},\dots,\bfx_{\beta_{i_n}}) .
\end{equation*}
Adding the above expressions we obtain the thesis:
\begin{equation*}
\begin{split}
   \pp(\bfx_{\alpha_1},\dots,\bfx_{\alpha_{M-h}}&\in S[1,T],\bfx_{\beta_1},\dots,\bfx_{\beta_{h}}\notin S[1,T])
   \\
   =&
\sum_{n=0}^{h}(-1)^n\sum_{\beta_{i_1}<\dots<\beta_{i_n}}
   f( 
   \bfx_{\alpha_1},\dots,\bfx_{\alpha_{M-h}},
\bfx_{\beta_{i_1}},\dots,\bfx_{\beta_{i_n}}) ,
   \end{split}
\end{equation*}
i.e., \eqref{marginals1}.

\end{proof}

    Consider now again a set of agents $V$, i.e., a list $\{S^1,\dots,S^N\}$  of $N$ independent simple random walks: by \eqref{full_Q} and Proposition \ref{prop_marginals}, the probability distribution of the intersection bipartite graph is completely determined by the set functions $Q^T_{\bfx_A}$ computed on all subsets of $F$.  In Subsection \ref{subsec_analytical results} we prove formulas to determine $Q^T_{\bfx_A}$ while in the next Subsection we focus on the special case when $A$ reduces to a single point.

\subsection{An explicit formula for the distribution of first arrival times at point $\bfx\in\zz^2$}


As a first step, we recall a well-known expression for the probability that a simple random walk is at a given site at a specific step $n$. Let $\bfx=(i,j)\in \zz^2$, and consider a simple random walk starting at the origin, and write 
\begin{equation*}
   p_\bfx(n)=\pp(S(n)=\bfx|S(0)=\bf0) 
\end{equation*}
for the probability that the random walk is at site $\bfx$ at time $n$. 
\begin{lemma}
Assume that $i,j\ge0$. Then 
\begin{equation} 
  p_\bfx(n)=
  \begin{cases}
\displaystyle    \frac1{4^n}   \binom{n}{(n+i-j)/2}\binom{n}{(n+i+j)/2} & 
   \text{ when } n=i+j \mod 2 \text{ and } n\ge i+j 
\\
\displaystyle 0 & \text{ when } n\ne i+j \mod 2\text{ or } n< i+j  .
   \end{cases}
   \label{probability_in_x}
\end{equation}
When $\bfx$ does not belong to the first quadrant $i$ and $j$ are replaced by $|i|$ and $|j|$.
\end{lemma}

Notice that \eqref{probability_in_x} is consistent with the fact that the simple random walk in $\zz^2$ is the superposition of two independent one-dimensional simple random walks along the directions forming angles of $\pm\pi/4$ with the coordinate axes. Of course, \eqref{probability_in_x} reduces to
\begin{equation*}
 p_\bfzero(n)=\frac1{4^{n}}\binom{n}{n/2}^2,\qquad n \text{  even},
\end{equation*}
with  $p_\bfzero(n)=0$ for $n$ odd.

We are now in position to obtain an explicit formula for the distribution of the first arrival time at a given site. 
\begin{proposition}
 Let $\bfx=(i,j)\in \zz^2$, and consider a simple random walk starting at the origin. Let \begin{equation*}
    q_\bfx(n)=\pp(\tau_\bfx=n),
\end{equation*}
the distribution of the first arrival time  $\tau_{\bfx}$ of the random walk at $\bfx$. Then $q_\bfx(n)$ is solution of
\begin{equation}
p_\bfx(n)=\sum_{k=1}^n q_\bfx(k)p_{\bfzero}(n-k).
\label{convolution_0}
\end{equation}

\end{proposition}

\begin{proof}
    We use the Markov property of the random walk to split its sample into two segments. The first accounts for reaching $\bfx$ at time $k, (k=1,2,...,n)$ and the second for the return to $\bfx$ in $(n-k)$ steps. We then observe that the probability of returning to $\bfx$, starting from $\bfx$ coincides with $p_{\bfzero}(n-k)$ thanks to the symmetry of the random walk.
\end{proof}
To solve \eqref{convolution_0}, we use an inversion formula for the discrete convolution (Lemma \ref{lemma_convolution} in Appendix A.1) alternative to the formula involving the determinant of Toeplitz matrices \cite{gould}, by treating the relation
\begin{equation}
    y_n=\sum_{k=1}^n x_kz_{n-k},\qquad n\in \nn.
    \label{convolution1_0}
\end{equation}
as a system of linear equations in the unknowns $x_i$, with $z_0=1$, and $y_i$, $z_i$ assigned. 
Let  $Z_\text{even}=\{\bfx=(i,j)\in \zz^2:i+j=0\mod 2\}$ and $Z_\text{odd}=\{\bfx=(i,j)\in \zz^2:i+j=1\mod 2\}$.

\begin{proposition} \label{prop_first_arrival_time} The distribution of first arrival times at $\bfx=(i,j)$ of a simple random walk in $\zz^2$ for $n\ge|i|+|j|$ is
\begin{equation}
 q_\bfx(n)=\pp(\tau_\bfx=n)
 =\begin{cases}
\displaystyle\sum_{k=0}^{m-1}a_{2k}p_\bfx(2m-2k)& n=2m,\qquad \bfx\in Z_\text{even},
\\
\displaystyle\sum_{k=0}^{m-1}a_{2k}p_\bfx(2m+1-2k)& n=2m+1,\bfx\in Z_\text{odd},
\\ 0&\text{otherwise,}
 \end{cases}
 \label{solution_first_arrival_time1}
\end{equation}
with 
\begin{equation}
a_0=1,\qquad a_{2k}=\sum_{k_1+\dots+k_j=k}(-1)^j p_\bfzero(2k_1)\dots p_\bfzero(2k_j).
\label{solution_first_arrival_time2}
\end{equation}
where the sum on the right-hand side of \eqref{solution_first_arrival_time2} is taken on all compositions of $k$, i.e., all ordered lists of integers whose sum is $k$, and $j$ is the length of the composition $k_1+k_2+\dots +k_j$.  When $n<|i|+|j|$ then $q_\bfx(n)=0$.

\end{proposition}

\begin{proof}
We use the inversion formula \eqref{solution_convolution_1} for the discrete convolution in Lemma \ref{lemma_convolution}. Consider first the term 
 \begin{equation*}
\sum_{h_1+\dots+h_j=h}(-1)^jz_{h_1}\dots z_{h_j}
=\sum_{h_1+\dots+h_j=h}(-1)^jp^\bfzero_{h_1}\dots p^\bfzero_{h_j},
\end{equation*}
in \eqref{solution_convolution_1}. Since $p_\bfzero(h_i)=0$ for $h_i$ odd, the above expression 
is non-zero only when $h_1=2k_1,\dots,h_n=2k_n$ and $h=2k$ are all even. Assume now that $\bfx\in Z_\text{even}$: then again $p_\bfx(n)=0$ for $n$ odd, and we obtain the first assertion of the thesis letting $n=2m$. The second assertion follows analogously for $n=2m+1$ odd.
\end{proof}

\begin{remark}    

 For instance, the first few terms for $ \bfx\in Z_\text{even}$ have the form
\begin{align*}
      q_\bfx(2)&=p_\bfx(2),\\
         q_\bfx(4)&=p_\bfx(4)-p_\bfzero(2)p_\bfx(2),\\
        q_\bfx(6)&=p_\bfx(6)-p_\bfzero(2)p_\bfx(4) +((p_\bfzero(2))^2-p_\bfzero(4))p_\bfx(2) ,\\
        q_\bfx(8)&=p_\bfx(8)-p_\bfzero(2)p_\bfx(6)+((p_\bfzero(2))^2-p_\bfzero(4))p_\bfx(4)
        +(-(p_\bfzero(2))^3+2p_\bfzero(2)p_\bfzero(4)-p_\bfzero(6))p_\bfx(2),
\end{align*}
and  for $ \bfx\in Z_\text{odd}$ 
\begin{align*}
      q_\bfx(1)&=p_\bfx(1),\\
         q_\bfx(3)&=p_\bfx(3)-p_\bfzero(2)p_\bfx(1)\\
        q_\bfx(5)&=p_\bfx(5)-p_\bfzero(2)p_\bfx(3) +((p_\bfzero(2))^2-p_\bfzero(4))p_\bfx(1) ,\\
        q_\bfx(7)&=p_\bfx(7)-p_\bfzero(2)p_\bfx(5) +((p_\bfzero(2))^2-p_\bfzero(4))p_\bfx(3)
        +(-(p_\bfzero(2))^3+2p_\bfzero(2)p_\bfzero(4)-p_\bfzero(6))p_\bfx(1).
\end{align*}
More explicitly, we list below the first few non-zero coefficients in \eqref{solution_first_arrival_time2}:
\begin{equation*}
  a_0=1,\quad   a_2=- 1/4,\quad a_4=-5/64 ,\quad  a_6=-11/256,\quad a_8=-121/4227,\quad a_{10}=-103/4895.
\end{equation*}   
\end{remark}

We now compute the probabilities  defined in \eqref{marginals}, which are the basic tools to compute the probability distribution of the intersection graphs. We start from a single point. 

\begin{proposition}\label{first_first_proposition}
Let $\bfx\in\zz^2$: then
\begin{equation}
Q_{\{\bfx\}}^\bfzero(T)= \pp(\tau_{\bfx}\le T).
 \label{first_arrival_time}
\end{equation}
Further, 
\begin{equation}
Q_{\{\bfx\}}^\bfzero(T)= \begin{cases}
\displaystyle
\sum_{h=1} ^mb_{m-h}p_\bfx(2h),
& \bfx\in Z_\text{even}, m=\lfloor T/2\rfloor
\\ \displaystyle
\sum_{h=1} ^mb_{m-h}p_\bfx(2h+1),
&\bfx\in Z_\text{odd}, m=\lfloor (T-1)/2\rfloor
\end{cases}
 \label{first_arrival_time2}
\end{equation}
where the $b_h$ are coefficients independent of $\bfx$ and T:
\begin{equation*}
b_h=\sum_{k=0}^{h}\sum_{k_1+\dots+k_j=k}(-1)^jp_{\bfzero}(2k_1)\dots p_{\bfzero}(2k_j).
\end{equation*}

\end{proposition}
\begin{proof}
The first result  follows from $Q_{\{\bfx\}}^\bfzero(T)=
 \pp(\bfx\in S[1,T])$. Now,  for $\bfx\in Z_\text{even}$, letting $m=\lfloor T/2\rfloor$ and recalling Proposition \ref{prop_first_arrival_time},
\begin{equation*}
\begin{split}
    \pp(\tau_{\bfx}\le T)=&\sum_{n=1} ^T \pp(\tau_{\bfx}=n)
    =\sum_{h=1} ^m q_{\bfx}(2h)
    =\sum_{h=1} ^m\sum_{k=0}^{h-1}a_{2k}p_\bfx(2h-2k)
    =\sum_{h=1}^m\sum_{k=1}^{h}a_{2h-2k}p_\bfx(2k)
    \\
=&
\sum_{k=1}^m\sum_{h=k}^{m}a_{2h-2k}p_\bfx(2k)   
=\sum_{h=1}^m\left(\sum_{k=0}^{m-h}a_{2k}\right)p_\bfx(2h)
=\sum_{h=1}^mb_{m-h}p_\bfx(2h),
\end{split}
\end{equation*}
where we have used a diagonal summation argument, $a_{2k}$ are defined in \eqref{solution_first_arrival_time2} and we have let $b_h=\sum_{k=0}^{h}a_{2k}$. An analogous argument holds for $Z_\text{odd}$. \end{proof}

\begin{remark}\label{Q_monotonicity}
Notice that, being a cumulative distribution, $Q_{\{\bfx\}}^\bfzero(T)$ is non-decreasing in $T$, which implies that, since $p_\bfx(h)\ge 0$, then also all the coefficients $b_h$ are non-negative.
\end{remark}

\subsection{The probability of hitting a given set of points}
\label{subsec_analytical results}
Given $\bfx_0 \in \zz^2$ and $\bfx_{A} =\{\bfx_{\alpha_1}, \bfx_{\alpha_2}, \dots, \bfx_{\alpha_k}\}\subset \zz^2$, we write  
\begin{equation}
 \tau^{\bfx_0 }_{\bfx_{A}} =\min\{n\in \nn :\bfx_{A}\subseteq S[1,n]\}
 \label{visit_time}
\end{equation}
for the first time the random walk has visited all elements of the set $\bfx_{A}$, 
starting from $\bfx_0 $.
If $k=1$, then $\bfx_{A} =\bfx_{\alpha_1}=\bfx$, and $\tau^{\bfx_0 }_{\bfx_{A}}=\tau^{\bfx_0 }_{\bfx}$ becomes the first arrival time of the random walk at $\bfx$, starting from $\bfx_0 $. 

We denote the distribution of \eqref{visit_time} by 
\begin{equation}
q_{\bfx_{A}}^{\bfx_0}(n)=\pp(\tau^{\bfx_0 }_{\bfx_{A}}=n),  
\label{prob_visit}
\end{equation}
i.e., the probability that the random walk completes the visit of all elements of the set $\bfx_{A}$ for the first time at time $n$ having started at $\bfx_0$ at time 0. Observe that when $k=1$ we get
$q_{\bfx_{A}}^{\bfx_0}(n)=q_{\bfx}^{\bfx_0}(n)=\pp(\tau^{\bfx_0 }_{\bfx}=n)$ is just the distribution of first arrival times at $\bfx$.

Recalling \eqref{marginals}, we write 
\begin{equation}
Q_{\bfx_{A}}^{\bfx_0}(n)=\pp(\tau^{\bfx_0}_{\bfx_{A}}\leq n)
=\sum_{m=1}^nq_{\bfx_{A}}^{\bfx_0}(m), 
\label{marginals_final}
\end{equation}
for the probability that the random walk completes the visit of all elements of the set $\bfx_{A}$ by time $n$ having started at $\bfx_0$ at time 0.

\begin{figure}[ht]
 	\begin{center}
 \includegraphics[width=.85\textwidth]{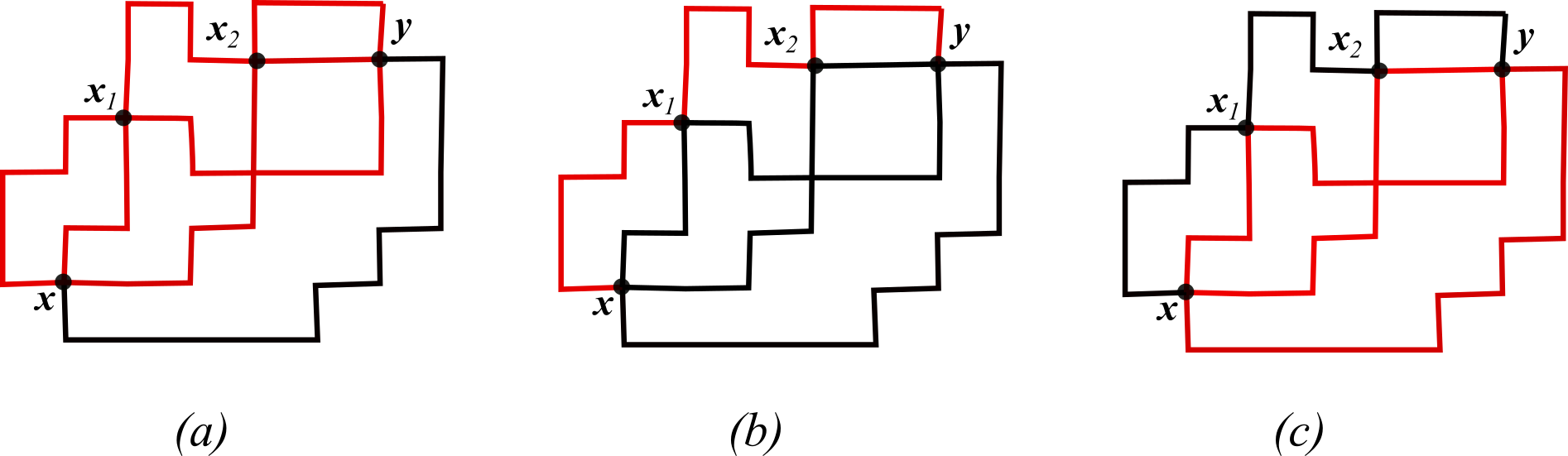}
 	\end{center}
 	\caption{\label{paths} Paths involved in the definitions of $g^{\bfx}_{\bfy} (t|\bfx_{A})$, $d^{\bfx}_{\bfy} (t|\bfx_{A})$ and $h^{\bfx}_{\bfy} (t|\bfx_{A})$, for $\bfx_{A}=\{\bfx_1,\bfx_2\}$; black paths are allowed, red paths are not.  (a) the quantity $g^{\bfx}_{\bfy} (t|\bfx_{A})$ only counts paths that do not meet any of the points in $A$; (b) the probability $d^{\bfx}_{\bfy} (t|\bfx_{A})$ only counts paths  that do not meet all points in $A$ before reaching $\bfy$ ; (c) the probability $h^{\bfx}_{\bfy} (t|\bfx_{A})$ only counts paths that meet all points in $A$ before reaching $\bfy$.
   }

 \end{figure}

We finally introduce the quantity
\begin{equation}
g_{\bfy}^{\bfx_0}(n|\bfx_{A})=\pp(\tau^{\bfx_0 }_{\bfy}=n|S[1,n]\cap \bfx_{A}=\emptyset), 
\label{not_hitting}
\end{equation}
which is the probability that the random walk, starting at $\bfx_0$, reaches point $\bfy$ for the first time at step $n$, without hitting any point in  the set $\bfx_{A}$ in the time interval $[0,n]$ (Figure \ref{paths}(a)). 
Our first result in this section is an iterative formula for the distribution of the visit times 
\eqref{prob_visit} and \eqref{marginals_final}.  Our approach here generalizes the procedure proposed in \cite{weiss1} for sets of two points.

\begin{theorem}
\label{general_rep_Q}
Given $\bfx_{A} =\{\bfx_{\alpha_1}, \bfx_{\alpha_2}, \dots, \bfx_{\alpha_k}\}\subset \zz^2$,  the distribution of the first visit times to $\bfx_{A}$ starting at $\bfx_0$ is
\begin{equation} 
q_{\bfx_{A}}^{\bfx_0}(m)=\sum_{u=1}^k \sum_{t=1}^{m} g^{\bfx_0}_{\bfx_{\alpha_u}} (t|\bfx_{A}\setminus\{\bfx_{\alpha_u}\}) q_{\bfx_{A}\smallsetminus\{\bfx_{\alpha_u}\}}^{\bfx_{\alpha_u}}(m-t)
\label{q_formula}
\end{equation}
while the probability that the whole set $\bfx_{A}$ has been visited by time $n$ is  
\begin{equation}
Q_{\bfx_{A}}^{\bfx_0}(n)=\sum_{m=1}^n q_{\bfx_{A}}^{\bfx_0}(m)
=\sum_{u=1}^k \sum_{t=1}^{n} g^{\bfx_0}_{\bfx_{\alpha_u}} (t|\bfx_{A}\setminus\{ \bfx_{\alpha_u}\}) Q_{\bfx_{A}\setminus\{\bfx_{\alpha_u}\}}^{\bfx_{\alpha_u}}(n-t).  
\label{Q_formula}
\end{equation}

\end{theorem}
\begin{proof}
Conditioning on the first point $\bfx_{\alpha_u}$, $u=1, \dots, k$ reached by the trajectory, we get \eqref{q_formula}. Relation \eqref{Q_formula} follows from \eqref{q_formula} by the usual diagonal summation identity $\sum_{m=1}^n\sum_{t=1}^m a_t b_{m-t}=\sum_{t=1}^n\sum_{m=t}^n a_t b_{m-t}=\sum_{t=1}^n\sum_{m=0}^{n-t} a_t b_{m}$.
\end{proof}

Hence, in order to compute the visit probabilities of the random walk, we can use the above iterative relation, which yields $Q_{\bfx_{A}}^{\bfx_0}(n)$ in terms of $Q_{\bfx_{A}\setminus\{\bfx_{\alpha_u}\}}^{\bfx_{\alpha_u}}(n)$, provided we can compute the probabilities $g_{\bfx}^{\bfx_0}(n|\bfx_{A})$ defined in \eqref {not_hitting}. The following result shows how to compute these objects.

\begin{theorem}
\label{T_general_rep_g}
Let $\bfx_{A} =\{\bfx_{\alpha_1}, \bfx_{\alpha_2}, \dots \bfx_{\alpha_k}\}$, $k\ge2$ and $\bfx_{\alpha_v}\in A$; then for each $n\in \nn$, 
\begin{equation}\label{g(n|x)}
g^{\bf0}_{\bfx_{\alpha_v}}(n|\bfx_{A}\smallsetminus \{\bfx_{\alpha_v}\})=q_{\bfx_{\alpha_v}}^{\bf0}(n) 
-\sum_{\substack{u=1\\ u\ne v}}^k \sum_{t=1}^n g^{\bf0}_{\bfx_{\alpha_u}} (t|\bfx_{A}\smallsetminus \{\bfx_{\alpha_u}\}) q^{\bfx_{\alpha_u}}_{\bfx_{\alpha_v}}(n-t),
\end{equation}
for $v=1,2,\dots, k$.
\end{theorem}

\begin{proof}
Notice first that, for $k=2$, \eqref{g(n|x)} coincides with (13) in \cite{weiss1}. Now,  choose $v=1$ for simplicity, and, for a general $k$, observe that \( q_{\bfx_{\alpha_1}}^{\bf0}(n) \), i.e. the probability of reaching \( \bfx_{\alpha_1} \) for the first time at step \( n \), can be expressed as the sum of two probabilities. The first is the probability that the random walk reaches \( \bfx_{\alpha_1} \) for the first time at step \( n \), starting from \( \bfx_0 \), before visiting any of the  points in \( \bfx_{A} \smallsetminus \{\bfx_{\alpha_1}\} = \{ \bfx_{\alpha_2}, \dots, \bfx_{\alpha_k} \} \). The second is the probability that the random walk visits at least one point in \( \bfx_{A} \smallsetminus \{\bfx_{\alpha_1}\} \)  by time \( n \), before reaching $\bfx_{\alpha_1}$.  

The first term corresponds to our unknown quantity, \( g^{\bf0}_{\bfx_{\alpha_1}}(n|\bfx_{A} \smallsetminus \{\bfx_{\alpha_1}\}) \). To express the second term, we condition on \( \bfx_{\alpha_u} \), the first element of the set \( \bfx_{A} \smallsetminus \{\bfx_{\alpha_1}\} \) that is reached. This approach allows us to divide the sample paths into two segments:  the first  consists of trajectories that start at \( \bf0 \) and reach \( \bfx_{\alpha_u} \) at time \( t \) for the first time without visiting any other point in \( \bfx_{A} \smallsetminus \{\bfx_{\alpha_u}\} \); the second segment consists of trajectories that, starting from \( \bfx_{\alpha_u} \), reach \( \bfx_{\alpha_1} \) for the first time in \( n-t \) steps.

The probability corresponding to the first segment is given by \( g^{\bf0}_{\bfx_{\alpha_u}}(t|\bfx_{A} \smallsetminus \{\bfx_{\alpha_u}\}) \), while the probability of the second segment is just $q^{\bfx_{\alpha_u}}_{\bfx_{\alpha_1}}(n-t) $.

\end{proof}

The probabilities  $g^{\bfx_0}_{\bfx}$ allow to compute a number of other interesting quantities related to the sites visited by a random walk, namely
\begin{itemize}
    \item $d_{\bfy}^{\bfx_0}(n|\bfx_{A})$, defined as  the probability that the random walk reaches $\bfy$ for the first time at step $n$ starting from $\bfx_0$ without having reached all elements of the set $\bfx_{A}$ by that step. In other words,  here we consider paths that may  hit some element of $A$, but \textit{not all} elements (Figure \ref{paths}(b)). Observe that  $\bfy$ can be an element of the set $\bfx_{A}$ and, if $\bfx_{A}=\varnothing$, we have  $g_{\bfy}^{\bfx_0}(n|\varnothing)=0$.
\item   $h_{\bfy}^{\bfx_0}(n|\bfx_{A})$,  defined as the probability that the random walk reaches $\bfy$ for the first time at step $n$ starting from $\bfx_0$, having reached all the elements of the set $\bfx_{A}$ by that time (Figure \ref{paths}(c)). 
\end{itemize}

\begin{theorem}
\label{T_general_rep_d}
Let $\bfx_{A} =\{\bfx_{\alpha_1}, \bfx_{\alpha_2}, \dots \bfx_{\alpha_k}\}$, $k=2,\dots$ and $\bfx_{\alpha_v}\in A$; then for each $n\in \nn$,

\begin{align}\label{d(n|x)}
d^{\bf0}_{\bfx_{\alpha_v}}&(n|\bfx_{A}\smallsetminus \{\bfx_{\alpha_v}\})=q_{\bfx_{\alpha_v}}^{\bf0}(n)\nonumber \\
&-\sum_{\substack{u=1\\ u\ne v}}^k \sum_{t=1}^n g^{\bf0}_{\bfx_{\alpha_u}} (t|\bfx_{A}\smallsetminus \{\bfx_{\alpha_u}\})\left[q^{\bfx_{\alpha_u}}_{\bfx_{\alpha_v}}(n-t)-d^{\bfx_{\alpha_u}}_{\bfx_{\alpha_v}}(n-t|\bfx_{A}\smallsetminus \{\bfx_{\alpha_u}, \bfx_{\alpha_v}\})\right]
\end{align}
for $v=1,2,\dots, k$.
\end{theorem}

\begin{proof}
The proof is similar to that of Theorem \ref{T_general_rep_g}. Without loos of generality, we  focus on the case $v=1$ but now split \( q_{\bfx_{\alpha_1}}^{\bf0}(n) \),  the probability of reaching the value \( \bfx_{\alpha_1} \) for the first time at step \( n \), as the sum of the probability that the random walk reaches \( \bfx_{\alpha_1} \) for the first time at step \( n \) having possibly visited some, but not all, of the points of  \( \bfx_{A} \smallsetminus \{\bfx_{\alpha_1}\}  \), and the probability that the random walk has visited every element of the set \( \bfx_{A} \smallsetminus \{\bfx_{\alpha_1}\} \) for the first time at step \( n \), regardless of their order.  As before, the first term corresponds to our unknown quantity, \( d^{\bf0}_{\bfx_{\alpha_1}}(n|\bfx_{A} \smallsetminus \{\bfx_{\alpha_1}\}) \) while, to compute the second term, we condition on \( \bfx_{\alpha_u} \), the first element of the set \( \bfx_{A} \smallsetminus \{\bfx_{\alpha_1}\} \) that is reached. As before this allows us to divide the sample paths into two segments. The first segment consists of trajectories that start at \( \bf0 \) and reach \( \bfx_{\alpha_u} \) at time \( t \) for the first time without having visited any point of \( \bfx_{A} \smallsetminus \{\bfx_{\alpha_u}\} \).  \\
The second segment consists of trajectories that, starting from \( \bfx_{\alpha_u} \), reach \( \bfx_{\alpha_1} \) in \( n-t \) steps while visiting all points of \( \bfx_{A} \smallsetminus \{\bfx_{\alpha_u}, \bfx_{\alpha_1}\} \).   While the probability corresponding to the first segment is given by \( g^{\bf0}_{\bfx_{\alpha_u}}(t|\bfx_{A} \smallsetminus \{\bfx_{\alpha_u}\}) \),  the probability of the second segment can be written as  
\[
\left[q^{\bfx_{\alpha_u}}_{\bfx_{\alpha_1}}(n-t) - d^{\bfx_{\alpha_u}}_{\bfx_{\alpha_1}}(n-t|\bfx_{A} \smallsetminus \{\bfx_{\alpha_u}, \bfx_{\alpha_1}\})\right].
\]

This expression arises from the observation that the relevant sample paths are those that visit all elements of the set \( \bfx_{A} \smallsetminus \{\bfx_{\alpha_u}, \bfx_{\alpha_1}\} \) for the first time.  Their probability can be computed as the difference between the probability of sample paths that reach for the first time \( \bfx_{\alpha_1} \) in \( n-t \) steps starting from \( \bfx_{\alpha_u} \) and the probability of those that do so without visiting all elements of the set \( \bfx_{A} \smallsetminus \{\bfx_{\alpha_u}, \bfx_{\alpha_1}\} \).  
\end{proof}

\begin{theorem}
\label{T_general_rep_h}
Let $\bfx_{A} =\{\bfx_{\alpha_1}, \bfx_{\alpha_2}, \dots \bfx_{\alpha_k}\}$, $k=1,2,\dots$ and $\bfx_{\alpha_v}\in A$: then, for each $n\in \nn$, 
\begin{align}\label{h(n|x)}
h^{\bf0}_{\bfx_{\alpha_v}}(n|\bfx_{A}\smallsetminus \{\bfx_{\alpha_v}\})=\sum_{\substack{u=1\\ u\ne v}}^k \sum_{t=1}^n g^{\bf0}_{\bfx_{\alpha_u}} (t|\bfx_{A}\smallsetminus \{\bfx_{\alpha_u}\})h^{\bfx_{\alpha_u}}_{\bfx_{\alpha_v}}(n-t|\bfx_{A}\smallsetminus \{\bfx_{\alpha_u},\bfx_{\alpha_v}\})
\end{align}
for each $v=1,2,\dots, k$.
\end{theorem}

\begin{proof}
By sequentially selecting a site in $\bfx_{A}$ and considering the probability of first arrival to that site without visiting all the elements of $\bfx_{A}$, and by applying the law of total probability, we obtain the thesis.
\end{proof}

\begin{remark}
Observe that Theorem \ref{T_general_rep_g} allows us to determine $g^{\bf0}_{\bfx_{\alpha_v}}(n|\bfx_{A}\smallsetminus \{\bfx_{\alpha_v}\})$  for $v=1, \dots, k$, as a solution of the linear system \eqref{g(n|x)}. Then, such solution allows to determine $d^{\bf0}_{\bfx_{\alpha_v}}(n|\bfx_{A}\smallsetminus \{\bfx_{\alpha_v}\})$ from Theorem \ref{T_general_rep_d}. Finally, the knowledge of $d^{\bf0}_{\bfx_{\alpha_v}}(n|\bfx_{A}\smallsetminus \{\bfx_{\alpha_v}\})$ determines $h^{\bf0}_{\bfx_{\alpha_v}}(n|\bfx_{A}\smallsetminus \{\bfx_{\alpha_v}\})$, using Theorem \ref{T_general_rep_h}.
\end{remark}

\section{An application to intersection graphs and hypergraphs}

As mentioned above, the analytical characterization of the intersection graph is difficult, one reason being that many bipartite graphs (hypergraphs) project to a given intersection graph. Therefore, we apply here the results of the previous sections to study a specific feature of this projection, namely, the fact that intersection graphs only describe pairwise associations between the agents, and these determine its cliques. However, $K$-cliques in the projection may as well result from the projection of $K$-faces of the intersection hypergraph, in which case the $K$ agents are associated  because they all have visited the same site.

To address this issue, we first compute the probability that a $K$-face is realized in the intersection hypergraph,  i.e., the probability that $K$ agents have visited the same site, possibly at different times.  

Recall that we denote by $F=\{\bfx_1,\dots,\bfx_\alpha,\dots,\bfx_M\}\subset \zz^2$ the sites and by $i,j\in \{1,\dots, N\}$ the agents, and $e_i$ and $e_j$ are the random subsets of $F$ corresponding to the sites visited by the random walks $S^i$ and $S^j$ up to step $T$. Also, we denote by $G=(V,E)$ the 2-section of the intersection hypergraph, with $V$ the set of agents and $E$ the set of edges of $G$. Our first result shows how to compute the probability that $K$ agents visit the same sites in $F$. 

\begin{proposition} 
\label{prop_intersection}
For $i_1,i_2,\dots,i_K\in V$ and $F=\{\bfx_1,\dots,\bfx_M\}$,  
\begin{equation}
\pp(e_{i_1}\cap \dots \cap e_{i_K}\ne\emptyset)=
\sum_{\emptyset \ne \bfx_A\subset F}(-1)^{|\bfx_A|+1}
[Q_{\bfx_A}(T)]^K,
\label{formula_for_intersection}
\end{equation}
where the sum is taken on all nonempty subsets of $F$ and $|\bfx_A|$ is the cardinality of $\bfx_A$.
\end{proposition}

\begin{proof}
Recall first that 
\begin{equation*}
\pp(e_{i_1}\cap \dots \cap e_{i_K}\ne\emptyset)=\pp(F\cap S^{i_1}[1,T]\cap \dots \cap S^{i_K}[1,T]\ne\emptyset).
\end{equation*}
Notice then that, since $F=\{\bfx_1\}\cup \dots\cup\{\bfx_M\}$,
\begin{equation*}
\begin{split}
\pp(e_{i_1}\cap \dots \cap e_{i_K}\ne\emptyset)=&
\pp(F\cap e_{i_1}\cap \dots \cap e_{i_K}\ne\emptyset)
\\
=&\pp((\{\bfx_1\}\cap e_{i_1}\cap \dots \cap e_{i_K})\cup
\dots 
 \cup (\{\bfx_M\}\cap e_{i_1}\cap \dots \cap e_{i_K}))\ne\emptyset)
\\
=&\pp(\bfx_1\in e_{i_1}\cap \dots \cap e_{i_K}\vee
\dots \vee \bfx_M\in e_{i_1}\cap \dots \cap e_{i_K})
\end{split}
\end{equation*}
and  by the inclusion/exclusion formula the last expression can be written as
\begin{equation*}
\sum_{h=1}^M(-1)^{h+1}
\sum_{1\le \beta_1<\dots<\beta_h\le M}
\pp(\{\bfx_{\beta_1},\dots\bfx_{\beta_h}\}\subseteq e_{i_1}\cap \dots \cap e_{i_K}),
\end{equation*}
which becomes 
\begin{equation*}
\sum_{h=1}^M(-1)^{h+1}
\sum_{1\le \beta_1<\dots<\beta_h\le M}
\pp(\{\bfx_{\beta_1},\dots\bfx_{\beta_h}\}\subseteq  S^{i_1}[0,T]\cap \dots \cap S^{i_K}[0,T]),
\end{equation*}
and, by independence of the random walks, this in turn can be written as
\begin{equation*}
\sum_{h=1}^M(-1)^{h+1}
\sum_{1\le \beta_1<\dots<\beta_h\le M}
[\pp(\{\bfx_{\beta_1},\dots\bfx_{\beta_h}\}\subseteq  S[0,T])]^K.
\end{equation*}
Recalling that $\pp(\{\bfx_{\beta_1},\dots\bfx_{\beta_h}\}\subseteq S[0,T])=Q_{\{\bfx_{\beta_1},\dots\bfx_{j_h}\}}(T)$, we obtain finally
\begin{equation*}
\sum_{h=1}^M(-1)^{h+1}
\sum_{1\le \beta_1<\dots<\beta_h\le M}
[Q_{\{\bfx_{\beta_1},\dots\bfx_{\beta_h}\}}(T)]^K.
\end{equation*}
Since the above sum is taken on all non-empty subsets of $F$, we obtain the thesis. 
\end{proof}

Notice that \eqref{formula_for_intersection} does not yield the probability that a specific $K$-face of the random intersection hypergraph is realized, which is given by \eqref{probability_alpha_face} instead.  Also, \eqref{formula_for_intersection} does not give the probability that $K$ agents are related by a $K$-face of the intersection hypergraph, because other agents may have visited the site. Neither this probability is given by  \eqref{probability_alpha_face}, because the agents may have visite more than one common site.  The actual probability that $K$ agents are related by a $K$-face of the intersection hypergraph is given by the result below.

\begin{proposition} [\bf The probability of a $K$- face of the intersection hypergraph.] Let $i_1,i_2,\dots,i_K\in V$ be a subset of $K$ agents. Then the probability that 
 these belong to a (possibly more than one) $K$-face of the intersection hypergraph is
\begin{equation}
\begin{split}
 \pp(e_{i_1}\cap \dots \cap e_{i_K}\ne\emptyset,&
(e_{i_1}\cap \dots \cap e_{i_K}) \cap
(e_{i_{K+1}}\cup \dots \cup e_{i_N})
=\emptyset
)
\\&=
\sum_{\emptyset \ne \bfx_A\subset F}(-1)^{|\bfx_A|+1}
Q_{\bfx_A}(T)^K(1-Q_{\bfx_A}(T))^{N-K},   
\end{split}
\label{formula_for_intersection2}
\end{equation}
where the sum is taken on all non-empty subsets of $F$ and and where $\{i_{K+1},\dots,i_{N}\}$ are the indices of agents in the complement of $\{i_1,i_2,\dots,i_K\}$ in $V$.
\end{proposition}

\begin{proof}
The proof follows the same steps as the proof of Proposition \ref{prop_intersection}. 
\end{proof}

Notice that both \eqref{formula_for_intersection} and \eqref{formula_for_intersection2} only depend on $F$ and the number of agents $K$. 

\subsection{The probability of an edge in the intersection  graph (one-mode projection). }
Let us turn now to the intersection graph $(G,V)$, which we view as either the one- mode projection of the intersection bipartite graphs, or the 2-section of the intersection hypergraph.
In order that  two agents be associated  in the intersection graph $G$ it is enough that they belong to some $K$-face of the intersection hypergraph. Hence,  the probability that an edge between two agents is realized in $G$ can be determined by \eqref{formula_for_intersection}:
\begin{equation}
\pp ((i,j) \in E)=\pp(e_i\cap e_j\ne\emptyset)=
\sum_{\emptyset \ne \bfx_A\subset F}(-1)^{|\bfx_A|+1}
[Q_{\bfx_A}(T)]^2.
\label{formula_for_intersectionG}
\end{equation}
Alternatively, we have the following equivalent characterization.
\begin{proposition}  For $i,j\in V=\{1,\dots,N\}$, 
\begin{equation}
\pp ((i,j) \in E)=\pp(e_i\cap e_j\ne\emptyset)= 1-\sum_{e'_i\subset F}\sum_{e'_j\subset F\setminus e'_i}\pp(e'_j)\pp(e'_i).  
\end{equation}
\end{proposition}

\begin{proof}
 Recalling \eqref{adjacency_agents}, 
\begin{equation*}
\pp ((i,j) \in E)= \pp(e_i\cap e_j\ne\emptyset)=1-\pp(e_i\cap e_j=\emptyset), 
\end{equation*}
and we have 
\begin{equation*}
\pp(e_i\cap e_j=\emptyset)=\sum_{e'_i\subset F}\pp(e_j\cap e'_i=\emptyset|e'_i)\pp(e'_i),
\end{equation*}
where the sum is on all subsets of $F$. On the other hand, 
\begin{equation*}
\pp(e_j\cap e'_i=\emptyset|e'_i)=\sum_{e'_j\subset F\setminus e'_i}\pp(e'_j), 
\end{equation*}
which yields the thesis.

\end{proof}

\begin{remark}
A first trivial consequence \eqref{formula_for_intersectionG} is that all edges are equiprobable in the intersection graph. However, these are non independent, i.e., in general, for $i,j,k\in (i,j)\in E$,
\begin{equation*}
\pp((i,j)\in E, (i,k)\in E)\ne
\pp((i,j)\in E)\pp ((i,k)\in E).
\end{equation*}
This can be seen already in a simple example. Let $F=\{\bfx_1,\bfx_2,\bfx_3\}$ with $\bfx_1=(2,1)$, $\bfx_2=(0,3)$, $\bfx_3=(-2,1)$, and $T=3$. Then, since no random walk of length 3 can hit more than one point in $F$, \eqref{formula_for_intersectionG} yields 
\begin{equation*}
\begin{split}
    \pp((i,j)\in E)=
\pp((i,k)\in E)=
\pp((j,k)\in E)
&=Q_{\{\bfx_1\}}(3)^2+Q_{\{\bfx_2\}}(3)^2+Q_{\{\bfx_3\}}(3)^2
\\
&=\left(\frac{3}{64}\right)^2+\left(\frac{1}{64}\right)^2+\left(\frac{3}{64}\right)^2.
\end{split}
\end{equation*}
On the other hand, by the same argument, 
\begin{equation*}
\begin{split}
\pp((i,j)\in E, (i,k)\in E)&=\pp(e_i\cap e_j\ne\emptyset,e_i\cap e_k\ne\emptyset )
\\
&=\pp(e_i\cap e_j\cap e_k\ne\emptyset)=Q_{\{\bfx_1\}}(3)^3+Q_{\{\bfx_2\}}(3)^3+Q_{\{\bfx_3\}}(3)^3
\\
&=\left(\frac{3}{64}\right)^3+\left(\frac{1}{64}\right)^3+\left(\frac{3}{64}\right)^3.
\end{split}
\end{equation*}
\end{remark}

\subsection{3-cliques vs. 3-faces}

The above results allow to work out completely a simple case that already suggests that 3-faces are much more likely than  3-cliques proper, a difference that, of course, is not visible in the one-mode projection of the social network. 

\begin{figure}[ht]
	\begin{center}
 \includegraphics[width=0.5\textwidth]{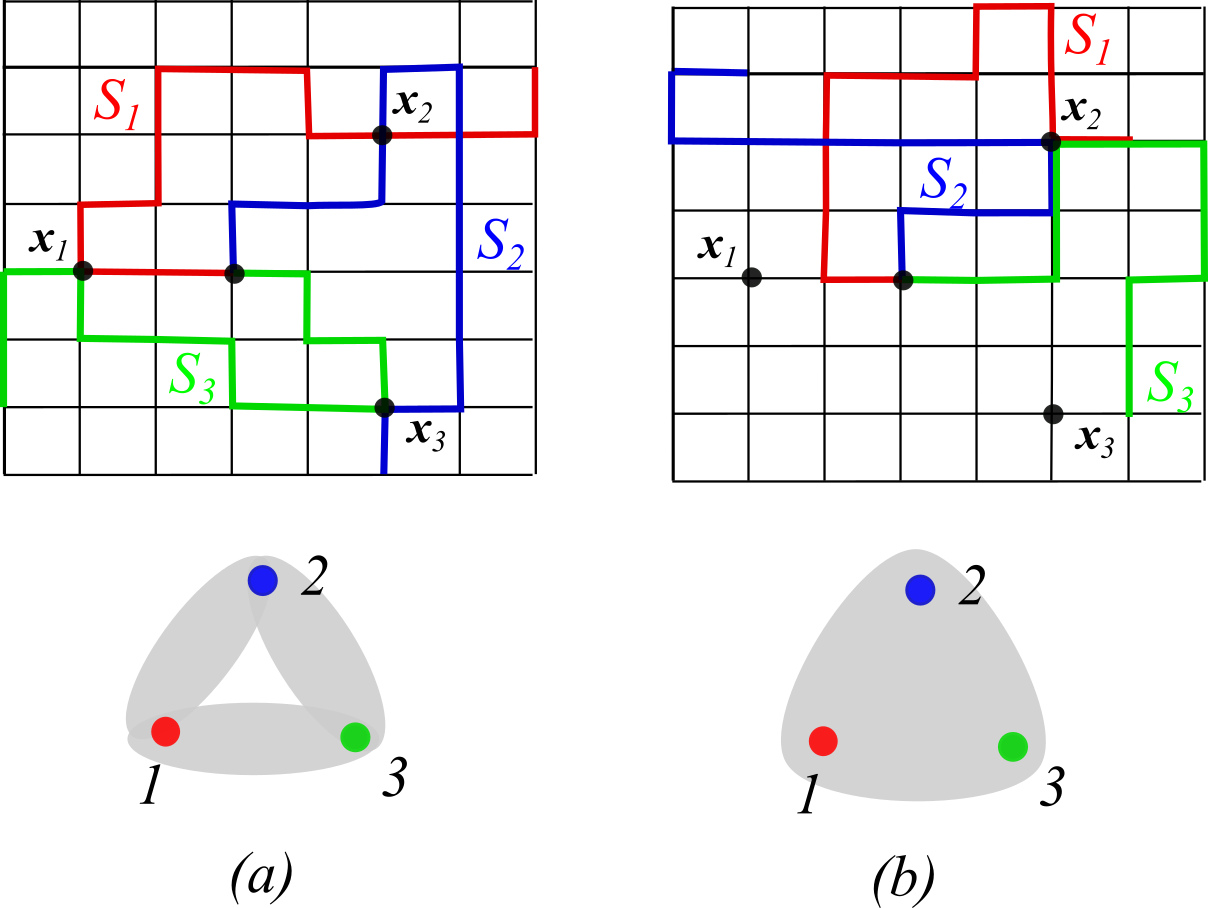}

	\end{center}
	\caption{\label{figurecliques} A sketch of two realizations of the random walks of three agents as discussed in Example \ref{examplecliqueversusface}. (a) the agents visit three different sites (top); the intersection hypergraph has three 2-faces (bottom); (b) the agents visit the same site (top);  the intersection hypergraph has a single 3-face (bottom).  Both realizations induce a 3-clique in the intersection graph.}
\label{figure_examplecliqueversusface}
\end{figure}

\begin{example}
\label{examplecliqueversusface}
Let $F=\{\bfx_1, \bfx_2,\bfx_3\}$, and $V=\{1,2,3\}$, so that there are only three sites and three agents (Figure \ref{figure_examplecliqueversusface}). In this simple case, \eqref{formula_for_intersection} yields the probability that a three-face is realized in the intersection hypergraph; omitting the argument $T$ in the notation, this probability is 
\begin{equation*}
    \pp(e_{1}\cap e_{2} \cap e_{3}\ne\emptyset)=
    Q_{\{\bfx_1\}}^3+Q_{\{\bfx_2\}}^3+Q_{\{\bfx_2\}}^3
    -Q_{\{\bfx_1,\bfx_2\}}^3-Q_{\{\bfx_1,\bfx_3\}}^3-Q_{\{\bfx_2,\bfx_3\}}^3
+Q_{\{\bfx_1,\bfx_2,\bfx_3\}}^3.
\end{equation*}
On the other hand, the probability that a 3-clique is realized in the intersection hypergraph is 
\begin{equation*}
\begin{split}
6\,\pp(\{\bfx_1,\bfx_2\})&\pp(\{\bfx_1,\bfx_3\})\pp(\{\bfx_2,\bfx_3\})
\\=&6\,(Q_{\{\bfx_1,\bfx_2\}}-Q_{\{\bfx_1,\bfx_2,\bfx_3\}})
(Q_{\{\bfx_1,\bfx_3\}}-Q_{\{\bfx_1,\bfx_2,\bfx_3\}})(Q_{\{\bfx_2,\bfx_3\}}-Q_{\{\bfx_1,\bfx_2,\bfx_3\}})
\\
=&6\,[Q_{\{\bfx_1,\bfx_2\}}Q_{\{\bfx_1,\bfx_3\}}Q_{\{\bfx_2,\bfx_3\}}
-Q_{\{\bfx_1,\bfx_2,\bfx_3\}}^3
\\
&
-Q_{\{\bfx_1,\bfx_2,\bfx_3\}}(Q_{\{\bfx_1,\bfx_2\}}Q_{\{\bfx_1,\bfx_3\}}
+
Q_{\{\bfx_1,\bfx_2\}}Q_{\{\bfx_2,\bfx_3\}}
+Q_{\{\bfx_2,\bfx_3\}}Q_{\{\bfx_1,\bfx_3\}})
\\
&+Q_{\{\bfx_1,\bfx_2,\bfx_3\}}^2(Q_{\{\bfx_1,\bfx_2\}}+Q_{\{\bfx_1,\bfx_3\}}+Q_{\{\bfx_2,\bfx_3\}})
].
\end{split}
\end{equation*}
Using the formulas in Section \ref{sec_analytical_results}, it is possible to compute both probabilities, The results are shown in Figure \ref{figure4_1} for some selected arrays of sites in the square lattice, and show that the probability that a 3-face is realized in the intersection hypergraph is an order of magnitude larger that the probability that three points are related by pairwise associations. 

\begin{figure}[t]
	\begin{center}
 \includegraphics[width=0.95\textwidth]{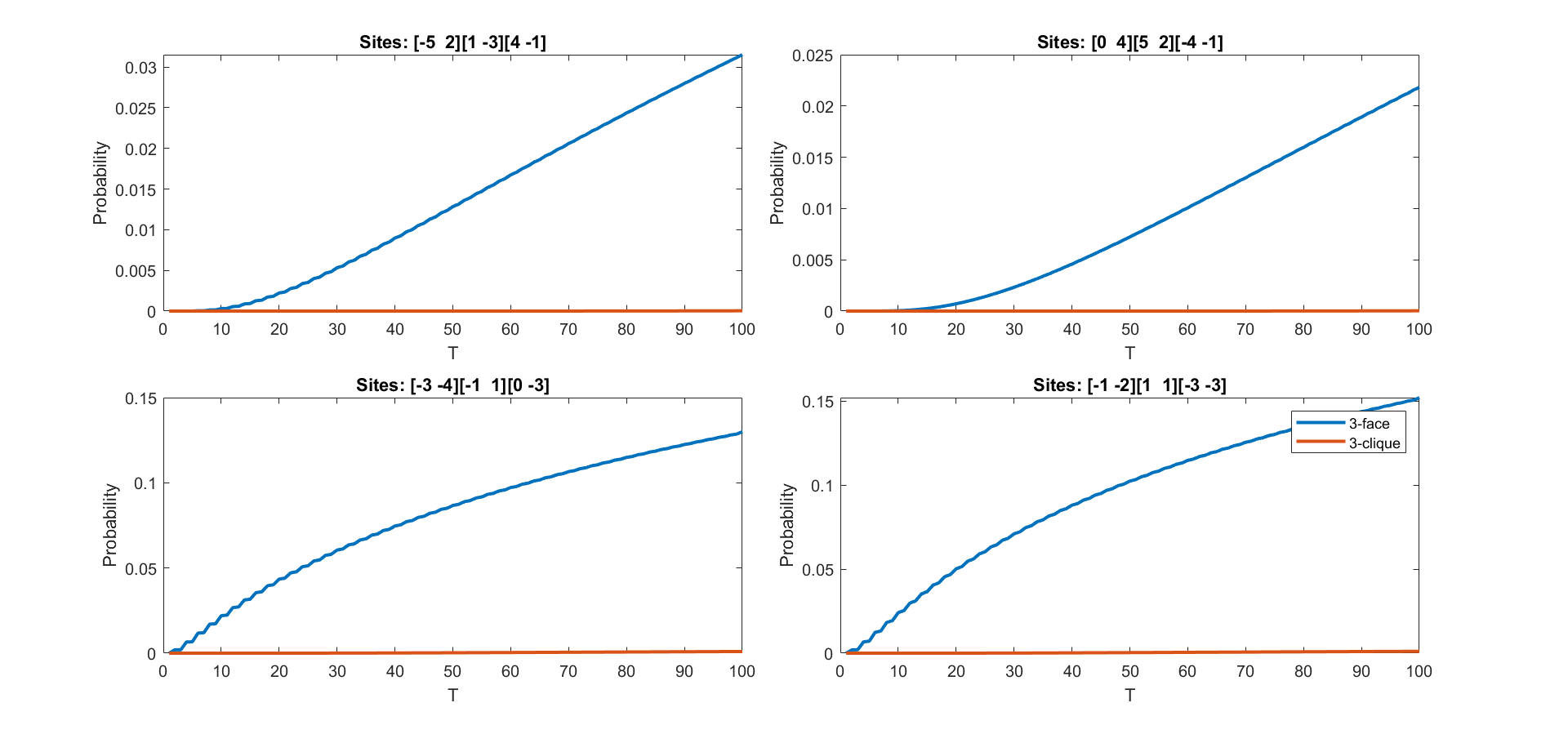}
  	\end{center}
	\caption{\label{figure4_1} The probability of a 3-face (blue) and of a 3-clique (red) as functions of the number of steps $T$ in the intersection hypergraph of  Example \ref{examplecliqueversusface} for 4 instances of randomly generated sites}

\end{figure}

\end{example}

\begin{figure}[ht]
	\begin{center}
 \includegraphics[width=0.35\textwidth]{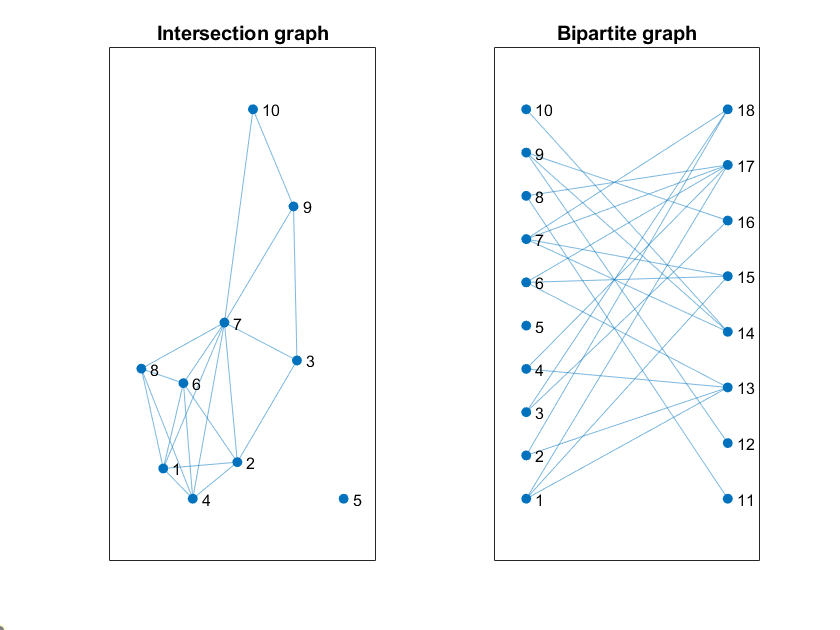}
  \includegraphics[width=0.35\textwidth]{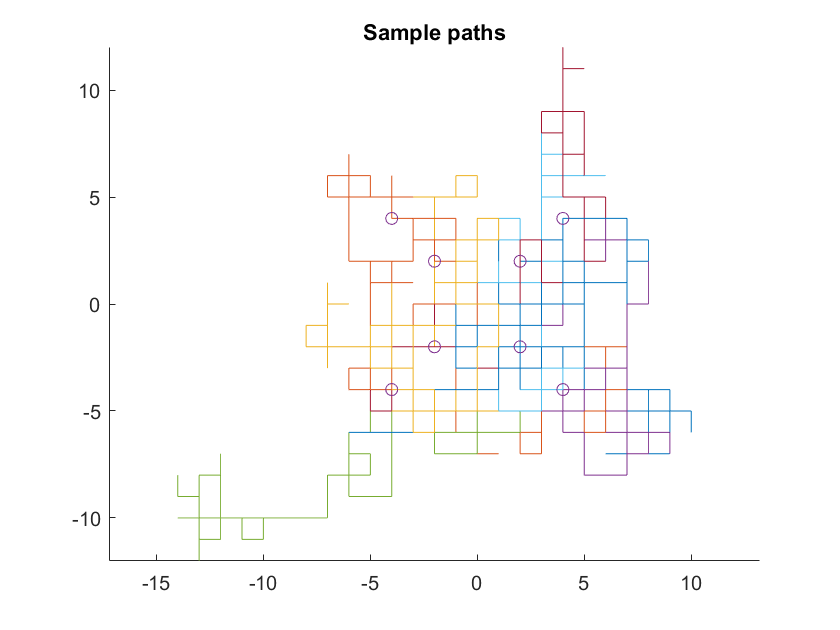}

     \includegraphics[width=0.35\textwidth]{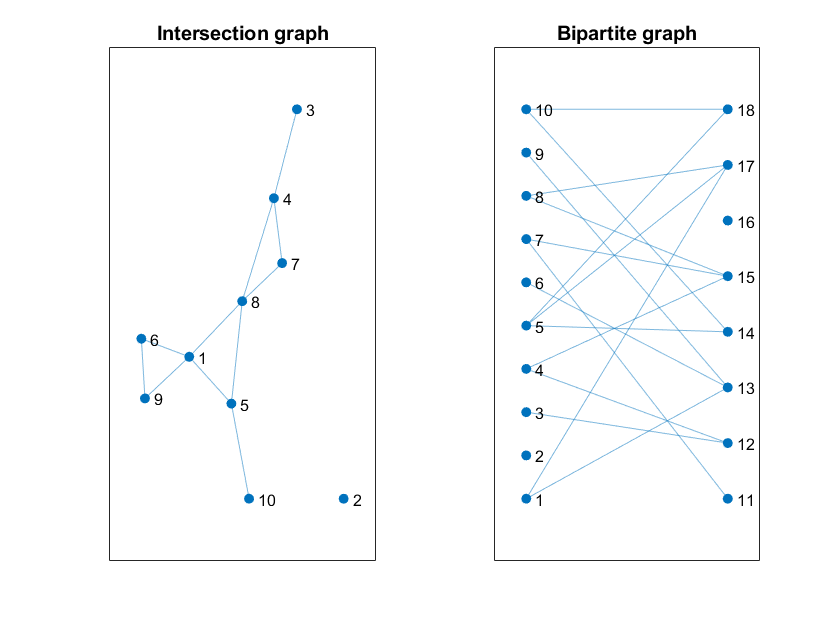}
  \includegraphics[width=0.35\textwidth]{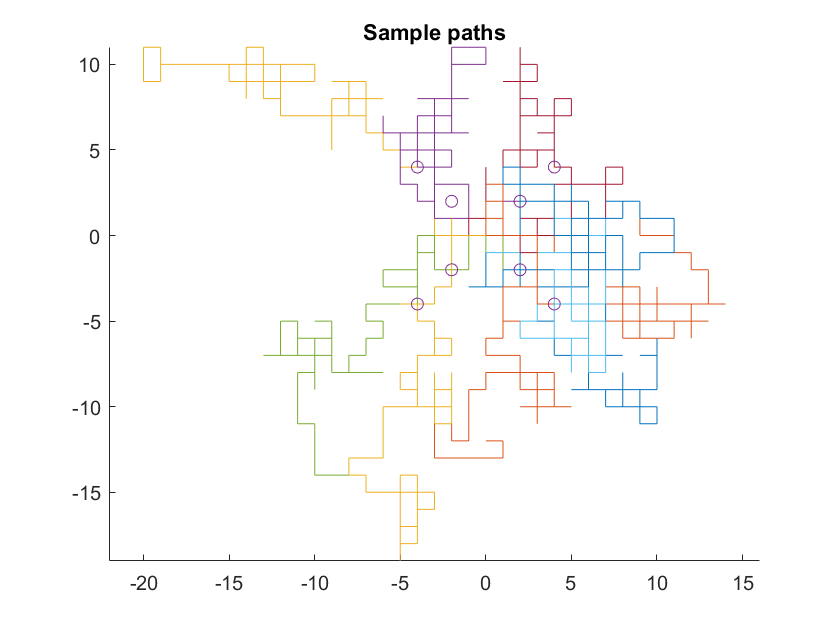}
   \includegraphics[width=0.35\textwidth]{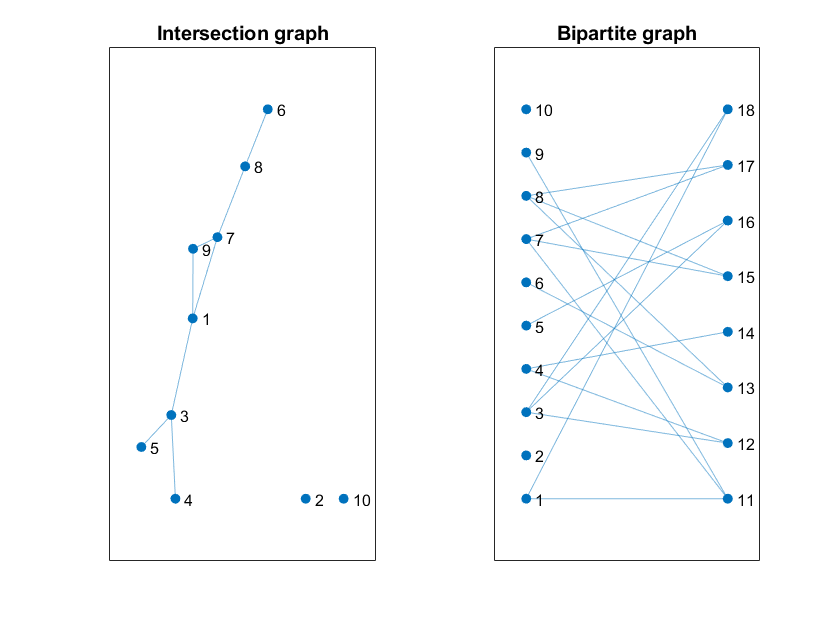}
  \includegraphics[width=0.35\textwidth]{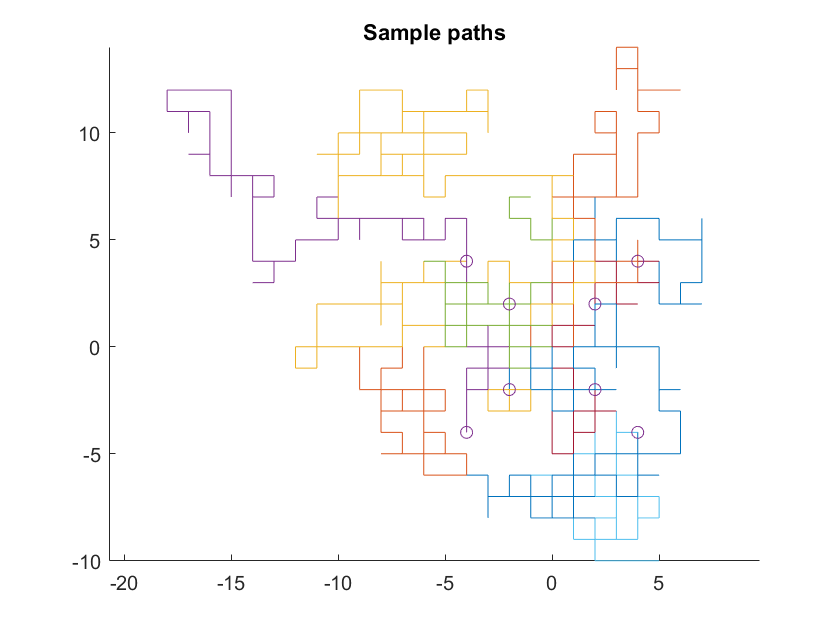}

	\end{center}
	\caption{\label{figure4} Some realizations of the bipartite intersection graph and its one-mode projection, together with a plot of the corresponding trajectories of the random walks.  Here $T=100$, there are 8 sites $F=\{(\pm2,\pm2),(\pm4,\pm4)\}$ and 10 agents. }

\end{figure}

We do not discuss here larger cliques. In fact, in order that $K$ agents be connected by a $K$-clique, they need to have $K(K-1)/2$ links. Now, each link corresponds to a site and, in order that this link is a maximal 2-face of the intersection hypergraph, each site must be visited by two agents only. Hence, each agent must visit at least $K-1$ distinct sites, and the probability of such event decreases as $K$ increases.

\section{Discussion} 
Association networks defined by the use of common resources play an important role in the study of animal behaviour. The interesting features of the resulting graphs are different from those characterizing  other social networks in that, in many cases,  the relevant topological features are cliques, and usual metrics such as degree, clustering, and so on, may be less relevant for the applications.  

For instance, in problems involving the transmission of parasites, each site can host an infection vector: in these cases it is important to understand how visits of the random walkers to common sites are related to the clique structure, while the number of visits to each site is less relevant.

However, cliques may arise in intersection graphs for different reasons: either pairs of agents visit different sites, or because many agents visit the same site (cf. Figure   \ref{figure_examplecliqueversusface}). One-mode projections cannot distinguish between these to situations, but from the point of view of applications, these are radically different, in that the nature of the association among the agents changes drastically.

Our point here is that this difference can be efficiently resolved using the hypergraph associated to the intersection bipartite graph: when a number $K$ of agents use the same resource, this yields a $K$-face of the hypergraph, which provides a more precise information on the relation between sites and cliques. 

To stress the importance of using hypergraphs in association networks, we have exploited the fact that it is possible to completely characterize the probability distribution of random hypergraphs defined by the intersection of the ranges of simple random walks. This allows to conclude that, when $K$ agents are mutually associated by the visits to common sites, i.e., a $K$- clique is present in the projected social network, it is much more likely that this is because they have visited a single common site, rather than different sites (cf. Example \ref{examplecliqueversusface} and Figure \ref{figure4_1}). 

This result is well-known in random intersection graphs, in which the probability that an object (here site) is related to an agent does not depend on the object and is independent of the choice of objects, in a special asymptotic regime when this probability is infinitesimal \cite{singer_cohen}. Here, however, the novelty is that, due to the nature of the random walk, visit to different sites are not independent, and the the visit probabilities are not infinitesimal. Hence, known results cannot be applied here, but a set of new  analytical formulas derived in Section \ref{sec_analytical_results} allow to evaluate precisely the association probabilities among the agents.  

To conclude, our work suggests that the abundant number of cliques observed in social networks characterized by the use of common resources is best described in terms of the hyperfaces of the corresponding intersection hypergraph, an approach that can be useful in many applications, an obvious example being the spread of infections in such networks.
Having highlighted some of the limits of the projection method, commonly used in this context, we suggest that data should be collected and analyzed using more sophisticated structures, such as hypergraphs, which are increasingly accepted as valuable tools in epidemiology.

\bmhead{Acknowledgements}

PC, LS and CZ thank the RILO 2023, RILO 2024 grant of the University of Turin.
LS and CZ were funded by the MIUR-PRIN 2022 project “Non-Markovian dynamics and non-local equations”, no. 202277N5H9 and by the Spoke 1 “FutureHPC \& BigData” of ICSC - Centro Nazionale di Ricerca in High-Performance-Computing, Big Data and Quantum Computing, funded by European Union - NextGenerationEU.
LS and CZ are also grateful to INdAM-GNAMPA.

\bmhead{Data availability statement}
The data used in Figure \ref{figure4_1} and \ref{figure4} were generated by MATLAB scripts based on the iterative formulas in Section \ref{sec_analytical_results}; the codes are available from the authors upon request. 

\begin{appendices}

\section{Discrete convolutions}\label{secA1}

We derive here an inversion formula for the discrete convolution, alternative to the formula involving the determinant of Toeplitz matrices \cite{gould}, by treating the relation
\begin{equation}
    y_n=\sum_{k=1}^n x_kz_{n-k},\qquad n\in \nn.
    \label{convolution1}
\end{equation}
as a system of linear equations in the unknowns $x_i$, with $z_0=1$, and $y_i$, $z_i$ assigned.

\begin{lemma}
\label{lemma_convolution}
The solution of \eqref{convolution1} is $x_1=y_1$ and 
\begin{equation}
    x_n=y_n +\sum_{h=1}^{n-1}\sum_{h_1+\dots+h_j=h}(-1)^jz_{h_1}\dots z_{h_j}y_{n-h},\qquad n>1,
    \label{solution_convolution_1}
\end{equation}
where the inner sum on the right-hand side is taken on all compositions of $h$, i.e., all ordered lists of integers whose sum is $h$, and $j$ is the length of the composition $h=h_1+h_2+\dots +h_j$. 
\end{lemma}

\begin{proof} We use induction on $n>1$. The assertion is clearly true for $n=2$. Assume that it is true for all $k\le n$, so that 
  \begin{equation*}
    x_k=y_k +\sum_{h=1}^{k-1}\sum_{h_1+\dots+h_j=h}(-1)^jz_{h_1}\dots z_{h_j}y_{k-h},\qquad 2\le k\le n.
\end{equation*}  
Then, by \eqref{convolution1}, 
  \begin{equation*}
    x_{n+1}=y_{n+1} -\sum_{k=1}^nz_{n+1-k}x_k,
\end{equation*}  
which becomes, recalling that $z_0=1$,
  \begin{align*}
    x_{n+1}=&y_{n+1} -\sum_{k=1}^nz_{n+1-k}
\left(y_k +\sum_{h=1}^{k-1}\sum_{h_1+\dots+h_j=h}(-1)^jz_{h_1}\dots z_{h_j}y_{k-h}\right)   
\\
=&y_{n+1} +\sum_{k=1}^nz_{n+1-k}
\left(\sum_{h=0}^{k-1}\sum_{h_1+\dots+h_j=h}(-1)^{j+1}z_{h_1}\dots z_{h_j}y_{k-h}\right)
\\
=&y_{n+1} +\sum_{k=1}^n\sum_{h=0}^{k-1}z_{n+1-k}a_hy_{k-h},
\end{align*}  
where we have written $a_h=\sum_{h_1+\dots+h_j=h}(-1)^{j+1}z_{h_1}\dots z_{h_j}$. Now, a diagonal summation argument shows that
\begin{equation*}
\sum_{k=1}^n\sum_{h=0}^{k-1}z_{n+1-k}a_hy_{k-h}
=\sum_{h=0}^{n-1}\sum_{k=h+1}^nz_{n+1-k}a_hy_{k-h}
=\sum_{h=0}^{n-1}\sum_{k=h+1}^{n}z_{k-h}a_hy_{n+1-k}
\sum_{k=1}^n\sum_{h=0}^{k-1}z_{k-h}a_hy_{n+1-k},
\end{equation*}   
from which it follows that 
 \begin{equation*}
    x_{n+1}=y_{n+1}  +\sum_{k=1}^n\left(\sum_{h=0}^{k-1}\sum_{h_1+\dots+h_j=h}
(-1)^{j+1}z_{k-h}z_{h_1}\dots z_{h_j}\right) y_{n+1-k},
\end{equation*}  
and the thesis is proved granted the validity of the identity 
\begin{equation*}
\sum_{h=0}^{k-1}\sum_{h_1+\dots+h_j=h}
(-1)^{j+1}z_{k-h}z_{h_1}\dots z_{h_j}=
\sum_{k_1+\dots+k_i=k}
(-1)^{i}z_{k_1}\dots z_{k_i}.
\end{equation*}  
This relation is true because all compositions of $k$ can be obtained by first taking all compositions of $1,2,\dots,h,\dots,k-1$  and completing them with $h$, which, by the way,  increases the length of each composition by one. 
\end{proof}
 For instance, the first few terms of \eqref{solution_convolution_1} have the form
\begin{align*}
        x_1&=y_1,\\
        x_2&=y_2-z_1y_1\\
        x_3&=y_3-z_1y_2 +(z_1^2-z_2)y_1\\
        x_4&=y_4-z_1y_3 +(z_1^2-z_2)y_2+(-z_1^3+2z_1z_2-z_3)y_1\\
        x_5&=y_5-z_1y_4 +(z_1^2-z_2)y_3+(-z_1^3+2z_1z_2-z_3)y_2
        +(z_1^4+z_2^2-3z_1^2z_2+2z_1z_3-z_4)y_1
        \end{align*}

\end{appendices}

\bibliography{biblio}

\end{document}